\documentclass[aps,prb,twocolumn,floatfix,floats,showpacs,superscriptaddress,raggedbottom]{revtex4-1}
\usepackage{graphicx,latexsym}
\usepackage{dcolumn}
\usepackage{amsmath}
\usepackage{amssymb,bm}
\usepackage[normalem]{ulem}

\def\mb#1{\mbox{\boldmath$#1$}}

\def\eq#1{Eq.\ (\ref{#1})}
\def\fig#1{Fig.\ \ref{#1}}

\usepackage{hyperref}
\hypersetup{
    pdfnewwindow=true,       
    colorlinks=true,         
    linkcolor=blue,          
    citecolor=blue,          
    filecolor=magenta,       
    urlcolor=black           
}

\begin{document}

\title{Electron transport through a quantum dot assisted by cavity photons}

\author{Nzar Rauf Abdullah}
\affiliation{Science Institute, University of Iceland,
        Dunhaga 3, IS-107 Reykjavik, Iceland}

\author{Chi-Shung Tang}
\email{cstang@nuu.edu.tw}
 \affiliation{Department of Mechanical Engineering,
        National United University, 1, Lienda, Miaoli 36003, Taiwan}

\author{Andrei Manolescu}
 \affiliation{Reykjavik University, School of Science and Engineering,
              Menntavegur 1, IS-101 Reykjavik, Iceland}

\author{Vidar Gudmundsson}
\email{vidar@raunvis.hi.is}
 \affiliation{Science Institute, University of Iceland,
        Dunhaga 3, IS-107 Reykjavik, Iceland}

%

\begin{abstract}
We investigate transient transport of electrons through a single-quantum-dot 
controlled by a plunger gate. The dot is embedded in a finite wire that is weakly 
coupled to leads and strongly coupled to a single cavity photon mode. 
A non-Markovian density-matrix formalism is employed to take into account 
the full electron-photon interaction in the transient regime.  
In the absence of a photon cavity, a resonant current peak can be
found by tuning the plunger gate voltage to lift a many-body state of the system into
the source-drain bias window. In the presence of an $x$-polarized photon field,
additional side peaks can be found due to photon-assisted transport.  By appropriately
tuning the plunger-gate voltage, the electrons in the left lead are allowed to make
coherent inelastic scattering to a two-photon state above the bias window if 
initially one photon was present in the cavity. However, this photon-assisted feature 
is suppressed in the case of a $y$-polarized photon field
due to the anisotropy of our system caused by its geometry.  
\end{abstract}

\pacs{73.23.-b, 42.50.Pq, 73.21.Hb, 78.20.Jq}


\maketitle

%
%

\section{Introduction}

Electronic transport through quantum dot (QD) related systems has received tremendous
attention in recent years due to its potential application in various fields, such as
implementation of quantum computing,\cite{Nakamura786.398} nanoelectromechanical
systems,\cite{Villavicencio92.192102} photodetectors,\cite{Kouwe97.113108} and biological
sensors.\cite{Jin.834139} The QD embedded structure can be fabricated in a
two-dimensional electron gas, controlled by a plunger-gate voltage, and connected to
the leads by applying an external source-drain bias voltage.

The electronic transport under the influence of time-varying external fields is one of
the interesting areas. The transport phenomena in the presence of photons have been
intensively studied in many mesoscopic
systems.~\cite{Pedersen14.12993,Stoof53.1050,Torres72.245339,Kouwenhoven50.2019,Niu56.R12752,Hu62.837,Tang127.292,
Watzel99.192101,Shibata109.077401,Ishibashi314.437}  Various quantum confined geometries to
characterize the photon-assisted features are for example a quantum ring with an
embedded dot for exploring mono-parametric quantum charge pumping,~\cite{Torres72.245339} a
single QD for investigating the single-electron (SE) tunneling,~\cite{Kouwenhoven50.2019}
a quantum wire for studying the electron population inversion,~\cite{Niu56.R12752} and a
quantum point contact involving photon-induced intersubband
transitions.~\cite{Hu62.837,Tang127.292} Recently, electrical properties of double QD
systems influenced by electromagnetic irradiation have been
studied,\cite{Watzel99.192101,Shibata109.077401} pointing out spin-filtering
effect,\cite{Watzel99.192101} and two types of photon-assisted tunneling related to the
ground state and excited state resonances.~\cite{Shibata109.077401} The classical and quantum
response was investigated experimentally in terms of the sharpness of the transition
rate which depends on the thermal broadening of the Fermi level in the electrodes and
the broadening of the confined levels.~\cite{Ishibashi314.437}

In the above mentioned examples the photon-assisted transport was induced by a classical
electromagnetic field. It is also interesting to investigate electronic transport through a
QD system influenced by quantized photon field. A single-photon source is an essential
building block for the manipulation of the quantum information coded by a quantum
state.\cite{Imamog72.210}  This issue has been considered by calculating resonant
current carried by negatively charged excitons through a double QD system confined in a
cavity,\cite{Joshi102.537} where resonant tunneling between two QDs is assisted by a
single photon. However, modeling of transient electronic transport through a QD in a
photon cavity is still in its infancy.

To study time-dependent transport phenomena in mesoscopic systems, a number of
approaches have been employed. In closed systems, the Jarzynski equation was derived by
defining the free-energy difference of the system between the initial and final
equilibrium state in terms of stochastic Liouville equation\cite{Mukamel90.170604} or
microscopic reversibility.~\cite{Monnai72.027102}  In open quantum systems where the system
is connected to electron reservoirs, the Jarzynski equation can be derived using a
master equation approach to investigate fluctuation theorems~\cite{Esposito73.046129} and
dissipative quantum dynamics.~\cite{Crooks77.034101}  In order to investigate interaction
effects on the transport behavior, several approaches have been proposed based on the
quantum master equation (QME) applied to a quantum measurement of a two-state
system,~\cite{Rammer70.115327} calculation of current noise spectrum,~\cite{Luo76.085325} and the
counting statistics of electron transfers through a double QD.~\cite{Welack77.195315}
The QME describes the evolution of the reduced density (RD) operator caused by the
Hamiltonian of the closed system in the presence of the electron or photon reservoirs.
Thus, the QME usually consists of two parts, a part describing the unitary evolution
of the closed system, and a dissipative part describing the influence of the 
reservoirs.~\cite{Lambropoulos63.455}

In an open current-carrying system weakly coupled to leads, the master equations within
the Markovian and wide-band approximations have been commonly derived and
used.\cite{Kampen2.2001,Harbola74.235309,Gurvitz53.15932} The coupling to electron or photon
reservoirs can be considered to be Markovian and the rotating wave approximation are
often used for the electron-photon coupling.\cite{Kampen2.2001} The QME may reduce to a
``birth and death master equation" for populations,\cite{Harbola74.235309} or modified
rate equations.~\cite{Gurvitz53.15932}  The energy dependence of the electron tunneling
rate or the memory effect in the system are usually neglected.

The non-Markovian density-matrix formalism with energy-dependent coupling elements
should be considered to study the full counting statistics for electronic transport
through interacting electron systems.~\cite{Braggio96.026805,Emary76.161404,Bednorz101.206803}  It
was noticed that the Markovian limit neglects coherent oscillations in the transient
regime, and the rate at which the steady state is reached does not agree with the
non-Markovian model.\cite{Vaz81.085315} The Markov approximation shows significantly longer
time to reach a steady state when the tunneling anisotropy is high, thus confirming its
applicability only in the long-time limit. To  investigate the transient transport, a
non-Markovian density-matrix formalism involving energy-dependent coupling elements
should be explicitly considered.\cite{Vidar11.113007}

The aim of this work is to investigate how the $x$- and $y$-polarized single photon
mode influence the ballistic transient electronic transport through a QD embedded in a
finite quantum wire in a uniform perpendicular magnetic field based on the
non-Markovian dynamics.  We explicitly build a transfer Hamiltonian that describes the
contact between the central quantum system and semi-infinite leads with a switching-on coupling
in a certain energy range. By controlling the plunger gate, we shall demonstrate robust
photon-assisted electronic transport features when the physical parameters of the
single-photon mode are appropriately tuned to cooperate with the electron-photon
coupling and the energy levels of the Coulomb interacting electron system.

The paper is organized as follows. In Sec.~\ref{Sec:II}, we model a QD with interacting electrons
embedded in a quantum wire coupled to a single-photon mode in a uniform magnetic
field, in which the full electron-photon coupling is
considered.  The transient dynamics is calculated using a generalized QME
based on a non-Markovian formalism.  Section \ref{Sec:III} demonstrates the numerical
results and transient transport properties of the plunger-gate controlled electron
system coupled to the single-photon mode with either $x$- or $y$-polarization.
Concluding remarks will be presented in Sec.~\ref{Sec:IV}.

\section{Model and Theory}\label{Sec:II}

\begin{figure}[htbq]
 \includegraphics[width=0.45\textwidth,angle=0]{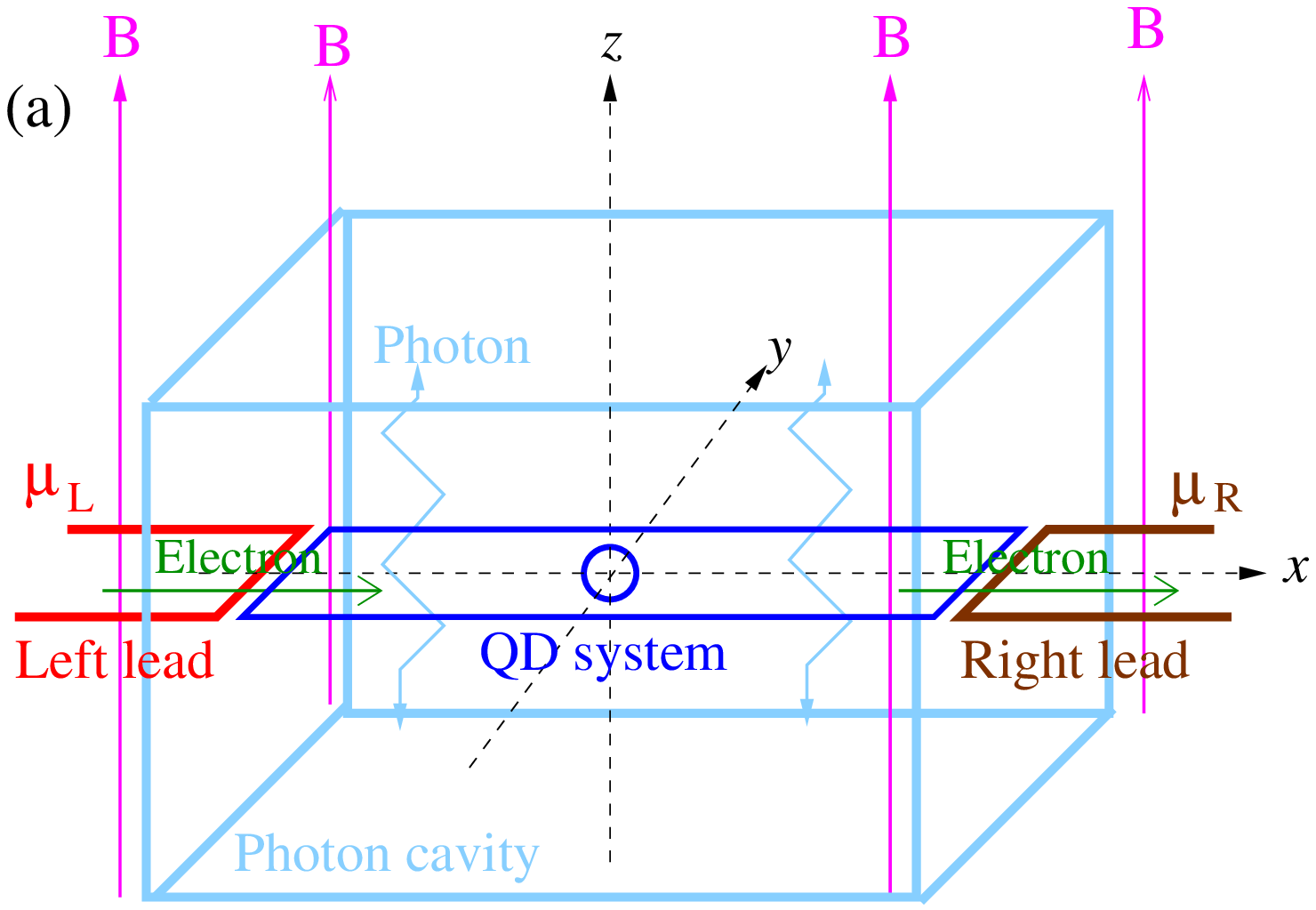}
 \includegraphics[width=0.48\textwidth,angle=0]{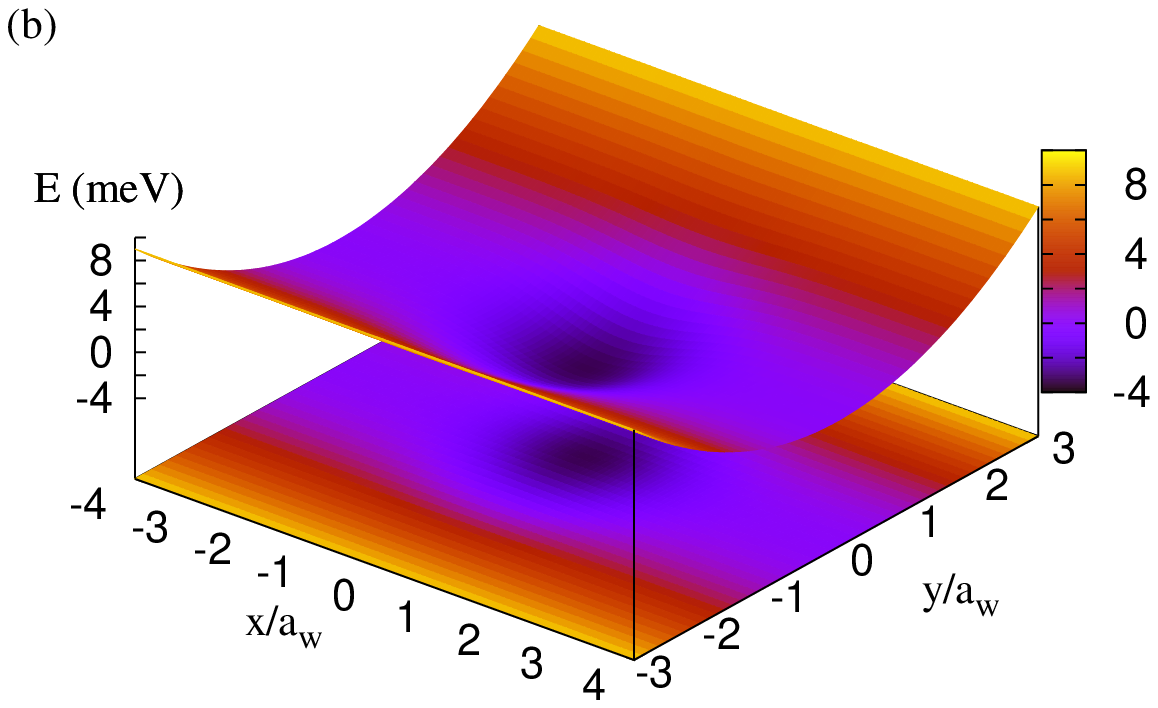}
 \caption{(Color online)
 (a) Schematic of a QD embedded in a quantum wire
     coupled to a photon cavity, connected to the left lead (red)
      with chemical potential $\mu_{\rm L}$, and the right lead (brown) with chemical potential $\mu_{\rm R}$
      in an external magnetic field $B$.
      (b) Schematic diagram depicts the potential representing the QD
      embedded in a quantum wire with parameters $B = 0.1~{\rm T}$,
      $a_{w} = 23.8~{\rm nm}$, and $\hbar \Omega_0 = 2.0~{\rm meV}$.}
      \label{QD_Cavity}
\end{figure}

In this section, we describe how the embedded QD, realized in a two-dimensional
electron gas in gallium arsenide (GaAs), can be described by the potential
$V_{\rm QD}$ in a finite quantum wire and its connection to the leads in a uniform
perpendicular magnetic field. The plunger-gate controlled central electronic system is
strongly coupled to a single photon mode that can be described by a many-body (MB)
system Hamiltonian $H_{\rm S}$, in which the electron-electron interaction and the
electron-photon coupling to the $x$- and $y$-polarized photon fields are explicitly
taken into account, as is depicted in \fig{QD_Cavity}(a). A generalized QME is
numerically solved to investigate the dynamical transient transport of electrons
through the single QD system.

\subsection{QD-embedded wire in magnetic field}

The electron system under investigation is a two-dimensional finite quantum wire that
is hard-wall confined at $x$ = $\pm L_{x}/2$ in the $x$-direction, and parabolically
confinement in the $y$-direction.  The system is exposed to an external perpendicular
magnetic field ${\bf B} = B\hat{z}$ defining a magnetic length $l = (h/eB)^{1/2} =
25.67 [B({\rm T})]^{-1/2}$~nm, and the effective confinement frequency $\Omega^2_w =
\omega^{2}_c + \Omega^2_0$ being expressed in the cyclotron frequency $\omega_c = e
B/m^*c$ as well as in the bare confinement energy $\hbar\Omega_0$ characterizing the
transverse electron confinement. The system is scaled by the effective magnetic length
$a_w = (\hbar /m^* \Omega_w)^{1/2}$. Figure \ref{QD_Cavity}(b) shows the embedded QD
subsystem scaled by $a_w$, where the QD potential is considered of a symmetric Gaussian
shape
\begin{equation}\label{potential}
      V_{\rm QD}(x,y) = V_0\; \exp{\left[ -\beta_{0}^2\left( x^2 + y^2
      \right)    \right]}
\end{equation}
with strength $V_0 = -3.3$ meV and $\beta_{0}$ = $3.0\times10^{-2}~{\rm nm^{-1}}$ such
that the radius of the QD is $R_{\rm QD} \approx 33.3~{\rm nm}$.

\subsection{Many-Body Model}
In this section, we describe how to build up the time-dependent Hamiltonian $H(t)$ of
an open system that couples the QD-embedded MB system to the leads. The Coulomb and
photon interacting electrons of the QD system are described by a MB system Hamiltonian
$H_{\rm S}$. In the closed electron-photon interacting system, the MB-space
$\{|\breve{\nu})\}$ is constructed from the tensor product of the electron-electron
interacting many-electron (ME) state basis $|\nu)$ and the eigenstates $|N\rangle$ of
the photon number operator $a^{\dag}a$, namely $|\breve{\nu}) = |\nu)\otimes
|N\rangle$.~\cite{Olafur86.046701} The Coulomb interacting ME states of the isolated system
are constructed from the SE states.\cite{Abdullah52.195325} The time-dependent Hamiltonian
describing the MB system coupled to the leads
\begin{equation}
      H(t) = H_\mathrm{S} + \sum_{l=\mathrm{L,R}} \left[ H_l + H_{\mathrm{T}l}(t)
      \right]
      \label{Ht}
\end{equation}
consists of a disconnected MB system Hamiltonian $H_\mathrm{S}$, and ME Hamiltonian of
the leads $H_l$ where the electron-electron interaction is neglected. In addition, $L$
and $R$ refer to the left and the right lead, respectively. Moreover, $H_{\mathrm{T}l}(t)$
is a time-dependent transfer Hamiltonian that describes the coupling between the QD
system and the leads.

The isolated QD system including the electron-electron and the photon-electron
interactions is governed by the MB system Hamiltonian
\begin{eqnarray}\label{HS}
      H_\textrm{S} &=&  \sum_{i,j} \langle\psi_i \lvert  \frac{\bm{\pi}^2}{2m^*}
        + V_\mathrm{QD} + eV_\mathrm{pg}  \lvert \psi_j\rangle  d_i^{\dagger} d_j
      \nonumber \\
      && + H_\textrm{e-e} + H_{\rm ph}+ H_{\rm Z}
\end{eqnarray}
where $|\psi\rangle$ is a SE state, $d^\dagger_{i}$ ($d_{j}$) are the electron creation
(annihilation) operators in the central system, and $H_{\rm ph} =
\hbar\omega_{\textrm{ph}} a^{\dagger}a$ is the photon Hamiltonian.  In addition,
$\bm{\pi} = \bm{\pi}_e + \frac{e}{c} \mathbf{A}_{\rm ph}$ where $\bm{\pi}_e=
p+\frac{e}{c}\mathbf{A}_{\mathrm{ext}}$ is composed of the momentum operator $p$ of the
electronic system and the vector potential $\mathbf{A}_{\mathrm{ext}}$ =  ($0,-By,0$)
represented in the Landau gauge. $H_{\rm Z}$ is the Zeeman energy $\pm \frac{1}{2}
g^\ast \mu_B B$, where $\mu_B$ is the Bohr magneton and $g^\ast$ the effective Lande
$g$-factor for the material.

In the Coulomb gauge, the photon vector potential can be represented as
\begin{equation}
      \mathbf{A_{\rm ph}} = A_{\rm ph}
       \left( a+a^{\dagger} \right)\mathbf{\hat{e}}\, ,
\end{equation}
if the wavelength of the cavity mode is much larger than  the size of the central
system. Herein, $A_{\rm ph}$ is the amplitude of the photon field. The electron-photon
coupling strength is thus defined by  $g_{\rm ph} = e A_{\rm ph} \Omega_wa_w/c$. In
addition, $\mathbf{\hat{e}} = (e_x,0)$ indicates the electric field is polarized
parallel to the transport direction in a TE$_{011}$ mode, and $\mathbf{\hat{e}} =
(0,e_y)$ denotes the electric field is polarized perpendicular to the transport
direction in a TE$_{101}$ mode. Moreover, we introduce the plunger gate voltage $V_{\rm
pg}$ to control the alignment of quantized energy levels in the QD system relative to
the electrochemical potentials in the leads.  In the second term of \eq{HS},
$\hbar\omega_{\mathrm{ph}}$ is the quantized photon energy, and $a^{\dagger}(a)$ are
the operators of photon creation (annihilation), respectively. The last term
$H_\textrm{e-e}$ describes the electron-electron interaction.

In a second quantized form, the isolated MB system Hamiltonian $H_{\rm S}$ can be
separated as
\begin{equation}
      H_\textrm{S} = H_\textrm{e} + H_\textrm{ph} + H_\textrm{e-ph} + H_\textrm{Z}\, .
 \label{HS-e-ph}
\end{equation}
The first part of $H_{\rm S}$ is the Coulomb interacting electron Hamiltonian
\begin{equation}
       H_\textrm{e} =\sum_i \left(E_i+eV_{\mathrm{pg}}\right)d_i^{\dagger}d_i  +
        \frac{1}{2}\sum_{ijrs}\langle V_{\textrm{Coul}}\rangle
        d_i^{\dagger}d_j^{\dagger}d_sd_r \, ,
\end{equation}
where $E_i$ is the energy of a SE state, $V_{\mathrm{pg}}$ is the electrostatic potential 
of the plunger gate, and
\begin{eqnarray}
      &&\langle V_{\mathrm{Coul}}\rangle = \langle ij|V_{\mathrm{Coul}}|rs\rangle   \nonumber \\
      &=& \int d\mathbf{r} d\mathbf{r^{\prime}} \psi^\mathrm{S}_{i} ({\bf r})^*
      \psi^\mathrm{S}_{j} ({\bf r}')^* V({\bf r} - {\bf r}')
      \psi^\mathrm{S}_r ({\bf r}') \psi^\mathrm{S}_s ({\bf r})
\end{eqnarray}
are the Coulomb matrix elements in the SE state basis with $\psi^\textrm{S}({\bf r})$
being the SE state wavefunctions and the Coulomb interaction potential $V({\bf r} -
{\bf r}') $.\cite{Abdullah52.195325}  The second part in \eq{HS-e-ph} is the photon Hamiltonian
$H_\textrm{ph} = \hbar\omega_{\textrm{ph}} \hat{N}_{\rm ph}$ with $\hat{N}_{\rm ph} =
a^{\dagger} a$ being the photon number operator. The third part in \eq{HS-e-ph} is the
electron-photon coupling Hamiltonian
\begin{eqnarray}
      H_\textrm{e-ph} &=& g_{\rm ph}\sum_{ij}d_i^{\dagger}d_j\; g_{ij}
      \left\{a + a^\dagger\right\} \nonumber \\
      &&+\frac{g_{\rm ph}^2}{\hbar\Omega_w} \sum_{i}d_i^{\dagger}d_i
      \left[  \hat{N}_{\rm ph} + \frac{1}{2}\left( a^\dagger a^\dagger + aa  + 1 \right)\right]
\end{eqnarray}
with the dimensionless electron-photon coupling factor $g_{ij}$.\cite{Vidar85.075306} An
exact diagonalization method is utilized solving the Coulomb interacting ME Hamiltonian
for the central system.\cite{Yannouleas70.2067} In order to couple the central system to
the leads connecting to the left (right) reservoir with chemical potential $\mu_L$
($\mu_R$), it is important to consider all MB states in the system and SE states in the
leads within an extended energy interval $[\mu_R-\Delta_R, \mu_L + \Delta_L]$ to
include all the relevant MB states involved in the dynamical transient transport.

The second term in \eq{Ht} is the noninteracting ME Hamiltonian in the lead $l$ given
by
\begin{equation}
   H_l = \int d{\mb{q}}\, \epsilon_l(\mb{q}) {c^\dagger_{{\mb{q}}l}}
   c_{{\mb{q}}l}
\end{equation}
where we combine the momentum of a state $q$  and its subband index $n_{yl}$ in lead $l$
into a single dummy index $\mb{q}=(n_{yl},q)$, we thus use $\int d\mb{q} \equiv
\sum_{n_y}\int dq$ to symbolically express the summation and integration for
simplicity.  In addition, ${c^\dagger_{{\mb{q}}l}}$ and $c_{{\mb{q}}l}$ are,
respectively, the electron creation and annihilation operators of the electron in the
lead $l$.

The system-lead coupling Hamiltonian is expressed as
\begin{equation}
      H_{{\rm T}l}(t)= \chi_l(t) \sum_{i}\int d{\mb{q}}\, \left[  {c^\dagger_{{\mb{q}}l}} T_{{\mb{q}i}l} d_i
      +   d^\dagger_i (T_{{i\mb{q}l}})^* c_{{\mb{q}}l}\right]
\end{equation}
where $ \chi_l(t) = 1 - 2\{\exp [\alpha_l (t-t_0)] + 1\}^{-1}$ is a time-dependent
switching function with a switching parameter $\alpha_l$, and
\begin{equation}
 T_{{\mb{q}i}l} =
 \int d\mathbf{r} d\mathbf{r^{\prime}} \psi_{\mb{q}l}(\mathbf{r}')^*
 g_{\mb{q}il} (\mathbf{r},{\bf r'}) \psi^\mathrm{S}_i({\bf r})
 \label{Tlqn}
\end{equation}
indicates the state-dependent coupling coefficients describing the electron transfer
between a SE state $|i\rangle$ in the central system and the extended state
$|\mb{q}\rangle$ in the leads, where $\psi_{\mb{q}l}(\mathbf{r})$ is the SE wave
function in the $l$ lead and $g_{\mb{q}il} ({\bf r},{\bf r'})$ denotes the coupling
function.\cite{Vidar11.113007}

\subsection{General Formalism of the Master Equation}

The time evolution of electrons in the QD-leads system satisfies the Liouville-von
Neumann (Lv-N) equation~\cite{Breuer2002,Esposito76.031132}
\begin{equation}
      i\hbar \dot{W}(t)= \left[H(t),W(t)\right]
\label{LN}
\end{equation}
in the MB-space, where the density operator of the total system is $W(t)$ with the
initial condition $W(t<t_0) = \rho_\mathrm{L}\rho_\mathrm{R}\rho_\mathrm{S}$. Electrons
in the lead $l$ in steady state before coupling to the central QD system are described
by of the grand canonical density operator\cite{Jinshuang128.1234703}
\begin{equation}
      \rho_l=\frac{e^{-\beta (H_l-\mu_l N_l)}}{{\rm Tr}_l \{e^{-\beta(H_l-\mu_l N_l)}\}}
\end{equation}
where $\mu_l$ denotes the chemical potential of the $l$ lead, $\beta = 1/k_{B}T_l$ is
the inverse thermal energy, and $N_l$ indicates the total number of electrons in the
$l$ lead.  The Lv-N equation (\ref{LN}) can be projected on the central system by
taking trace over the Hilbert space of the leads to obtain the RD operator
$\rho(t)={\rm Tr}_\mathrm{L}{\rm Tr}_\mathrm{R} W(t)$ where $\rho(t_0) =
\rho_\mathrm{S}$.\cite{Haake3.1723,Haake1973}

We diagonalize the electron-photon coupled MB system Hamiltonian $H_{\rm S}$ within a
truncated fock-space built from 22 SE states $\{ |\mu\rangle
\}$,\cite{Vidar85.075306,Vidar61.305} and then the system is connected to the leads at time
$t=t_0$ thus containing a variable number of electrons.  We include all sectors of the
MB Fock space, where the ME states with zero to 4 electrons are dynamically coupled to
the photon cavity with zero to 16 photons. The diagonalization brings us a new
interacting MB state basis $\{|\breve{\nu})\}$, in which $|\breve{\nu}) = \sum_{\alpha}
{\cal W}_{\mu\alpha}|\breve{\alpha}\rangle$ with ${\cal W}_{\mu\alpha}$ being a unitary
transformation matrix with size $N_{\mathrm{MB}}\times N_{\mathrm{MB}}$.  SE states are
labeled with Latin indices and many-particle states have a Greek index.  The spin
information is implicit in the index.  The spin degree of freedom is essential to
describe correctly the structure of the few-body Fermi system. This allows us to obtain
the RD operator in the interacting MB state basis $\breve{\rho}(t) = {\cal
W}^\dagger\rho(t){\cal W}$.

Using the notation
\begin{eqnarray}
 \Omega_{\mb{q}l}(t) &=& U_\mathrm{S}^\dagger (t)
 \int_{t_0}^tds\:\chi_l(s)\Pi_{\mb{q}l}(s) \nonumber \\
  &&\times \exp{\left[ -\frac{i}{\hbar}(t-s) \epsilon_l(\mb{q})  \right]}
  U_\mathrm{S}(t)\, ,
\end{eqnarray}
where
\begin{eqnarray*}
 \Pi_{\mb{q}l}(s) &=& U_\mathrm{S}(s)
      \left[ \left(\breve{\cal T}_l\right)^{\dagger} \breve{\rho}(s)\left[ 1 -
      f_l\left( \epsilon(\mb{q})\right)
      \right] \right.  \\
      && - \left. \breve{\rho}(s)\left(\breve{\cal T}_l\right)^{\dagger}
      f_l\left( \epsilon(\mb{q}) \right)
      \right]
      U_\mathrm{S}^\dagger(s),
\end{eqnarray*}
and $U_\mathrm{S}(t)$ = $\exp[iH_\mathrm{S}(t-t_0) / \hbar]$ is the time evolution
operator of the closed central system, $f_l\left( \epsilon(\mb{q}) \right) =
\{\exp[\epsilon(\mb{q})-\mu_l]+1\}^{-1}$ is the Fermi function in the $l$ lead at
$t=t_0$, the time evolution of the RD operator can then be expressed as
\begin{eqnarray}\label{TGQME}
      \frac{d{\breve{\rho}}(t)}{dt}
      &=& -\frac{i}{\hbar}\left[ H_{\rm S},\breve{\rho}(t)\right]\\
      && -\frac{1}{\hbar^2}\sum_{l=L,R}\chi_l(t) \int d\mb{q}\: \left( \left[\breve{\cal
      T}_l(\mb{q}),\Omega_{\mb{q}l}(t)\right] + {\rm h.c.}\right).
\nonumber
\end{eqnarray}
The first term governs the time evolution of the disconnected central interacting MB
system. The second term describes  the energy dissipation of interacting electrons
through charging and discharging effects in the central system by the leads.  In the
second term, $\breve{{\cal T}}_l(\mb{q})$ is the interacting MB coupling matrix
\begin{equation}
      \breve{\cal T}_l(\mb{q})
      =\sum_{\mu,\nu}\breve{\cal T}_{\mu\nu l}(\mb{q})
      |{\bf \breve{\nu}})({\bf \breve{\mu}}|,
\label{Toperator}
\end{equation}
in which both the Coulomb interaction and the electron-photon coupling have been
included.  Here $\breve{\cal T}_{\mu\nu l}(\mb{q}) = \sum_i T_{i\mb{q}l}(\breve{\mu}
|d_i^{\dagger}|\breve{\nu})$ indicates the coupling of MB states $|\breve{\nu})$ in the
central system caused by the coupling to the SE states in the leads described by the
coupling matrix $T_{i\mb{q}l}$.

\subsection{Charge and Current}

We now focus on the physical observables that we calculate for the QD system. The mean
photon number in each MB state $|\breve{\nu})$ can be written as
\begin{equation}
      N_{\rm ph} = \left( \breve{\nu} \left| \hat{N}_{\rm ph} \right| \breve{\nu}
      \right),
\end{equation}
where $\hat{N}_{\rm ph}$ is the photon number operator. The average of the electron
number operator can be found by taking trace of the MB states $\{ |\breve{\nu} ) \}$ in
the Fock space, namely $\langle \hat{N}_{\rm e}(t)\rangle = {\rm Tr} \{ W(t)
\hat{N}_{\rm e} \}$.

The mean value of the interacting ME charge distribution in the QD system is thus
defined by
\begin{equation}
 \mathcal Q({\bf r},t) = e \sum_{i,j} \psi^*_{i}({\bf r}) \psi_j({\bf r})
 \sum_{\mu,\nu} (\breve{\mu} | d_i^{\dagger} d_j | \breve{\nu}) \breve{\rho}_{\nu\mu}(t)
      \label{Qxy}
\end{equation}
where $e>0$ stands for the magnitude of electron charge, and $\breve{\rho}_{\nu
\mu}(t)$ = $( \breve{\nu} |\breve{\rho}(t) | \breve{\mu} )$ is the time-dependent RD
matrix in the MB space.

In order to analyze the transient transport dynamics, we define the net charging
current
\begin{equation}
 I_Q(t) = I_L(t) + I_R(t)
 \end{equation}
where $I_L(t)$ indicates the partial charging current from the left lead into the
system, and $I_R(t)$ represents the partial charging current from the right lead into
the system. Here, the left and right partial currents $I_l(t)$ can be explicitly
expressed in the following form
\begin{eqnarray}
 I_l(t) &=& - \frac{e}{\hbar^2}\chi_l(t) \sum_{\mu} \int d\mb{q}\: \left( \breve{\mu} \left|
\left[\breve{\cal T}_l(\mb{q}),
 \Omega_{\mb{q}l}(t)\right] + {\rm h.c.}\right| \breve{\mu} \right)\, .\nonumber \\
 &&
 \label{Il}
\end{eqnarray}

\section{Results and Discussion}\label{Sec:III}

In this section, we consider a QD embedded in a finite quantum wire system, made of
high-mobility GaAs/AlGaAs heterostructure with electron effective mass $m^*=0.067m_e$
and relative dielectric constant $\varepsilon_r = 12.4$, with length $L_x=300$~nm and
bare transverse electron confinement energy $\hbar\Omega_0 = 2.0$~meV. A uniform
perpendicular magnetic field $B=0.1$~T is applied and, hence, the effective magnetic
length is $a_{w} = 23.8~{\rm nm}$, and the characteristic Coulomb energy is $E_{\rm C}
= e^2/(2\varepsilon_r a_w) \approx 2.44~{\rm meV}$.  The effective Lande $g$-factor
$g^\ast$ = 0.44.

We select $\beta_{0}$ = $3.0\times10^{-2}~{\rm nm^{-1}}$  such that the radius of the
embedded QD is $R_{\rm QD}$ = $1.4 a_w$.  The QD system is transiently coupled to the
leads in the $x$ direction that is described by the switching parameter $\alpha^l =
0.3$~${\rm ps}^{-1}$, and the nonlocal system-lead coupling strength $\Gamma_l =
1.58$~meV$\cdot$nm$^2$.\cite{Abdullah52.195325} A source-drain bias $V_{\rm bias}$ is applied,
giving rise to the chemical potential difference $\Delta\mu = eV_{\rm bias} = 0.1$~meV.

To take into account all the relevant MB states, an energy window
$\Delta_E = 5.5$~meV is considered to include all active states in the central system
contributing to the transport.  The temperature of the system is assumed to be
$T=0.01$~K such that the typical MB energy level spacing is greater than the thermal
energy, namely $\Delta E_{\rm MB} > k_B T$, the thermal smearing effect is thus
sufficiently suppressed. In the following, we shall select the energy $\hbar
\omega_{\rm ph}$ of the photon mode to be smaller than the characteristic Coulomb
energy, namely $E_{\rm C}
> \hbar \omega_{\rm ph}$.  In the following, we shall demonstrate the plunger-gate
controlled transient transport properties both in the case without a photon cavity and in the
case including a photon cavity with either $x$- or $y$-polarized photon field.

\subsection{Without photon cavity}

\begin{figure}[tbhq]
 \includegraphics[width=0.23\textwidth,height=0.26\textheight,angle=0,viewport=0 3 200 260,clip]{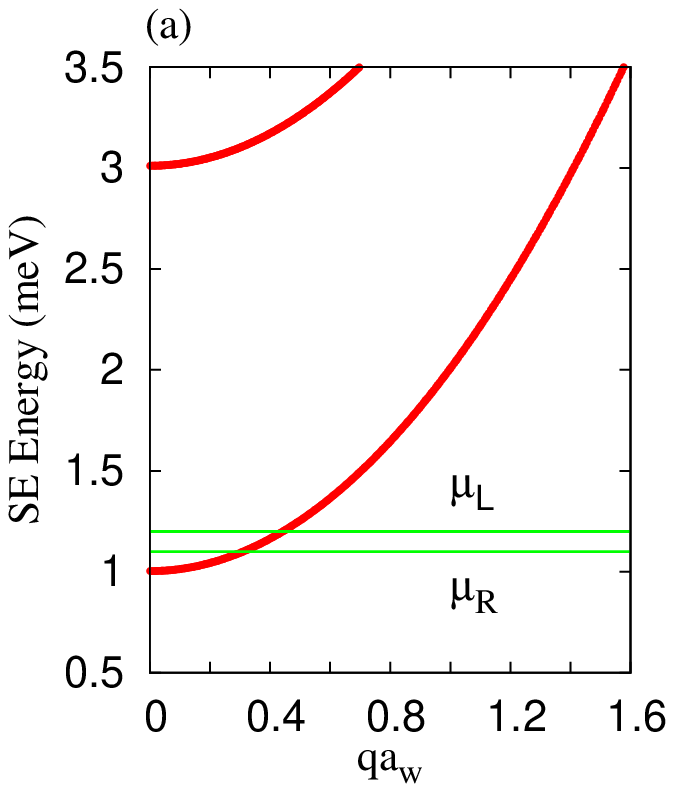}
 \includegraphics[width=0.23\textwidth,height=0.24\textheight,angle=0,viewport=0 3 200 240,clip]{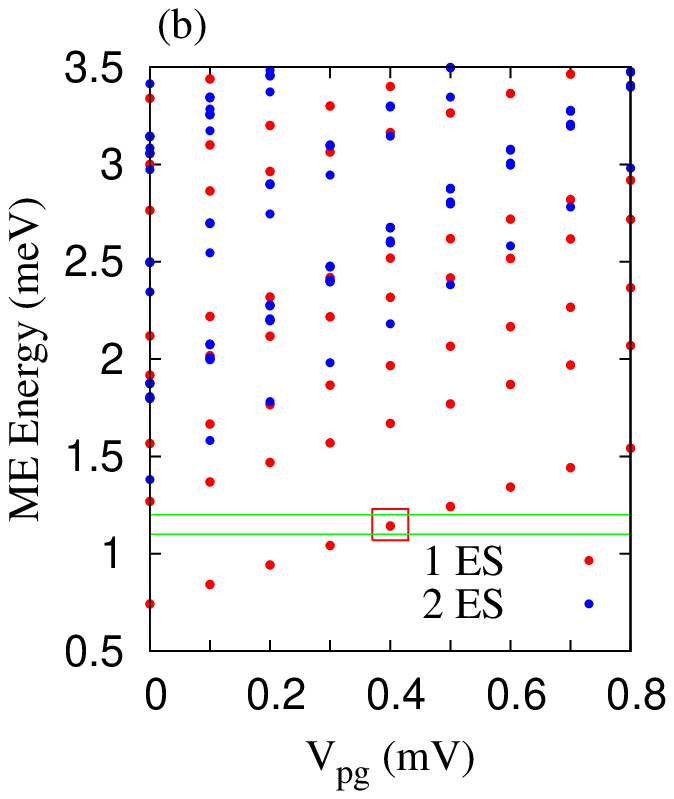}
 \caption{(Color online)
Energy spectra in the case of no photon cavity with magnetic field $B=0.1$~T. (a) SE
energy spectrum in the leads (red) is plotted as a function of wave number $q$, where
the chemical potentials are $\mu_L = 1.2\ {\rm meV}$ and $\mu_R = 1.1\ {\rm meV}$
(green). (b) ME energy spectrum in the central system as a function of plunger gate
voltage $V_{\rm pg}$ including SE states  (1ES, red dots) and two electron states (2ES,
blue dots). The SE state in the bias window is almost doubly degenerate due to the
small Zeeman energy.} \label{MBE_WOph}
\end{figure}

First, we consider the QD embedded in a quantum wire without a photon cavity in a
uniform magnetic field $B = 0.1$~T that is coupled to the leads acting as SE reservoirs
controlled by a source-drain bias. In \fig{MBE_WOph}(a), we show the SE energy spectrum
in the leads (red) as a function of wave number $q$ scaled by the effective magnetic
length $a_w^{-1}$. The first subband, $n_y = 0$, contributes to the propagating modes,
while higher subbands contribute to the evanescent modes. In addition, the chemical
potential (green) is $\mu_L = 1.2\ {\rm meV}$ in the left lead and $\mu_R = 1.1\ {\rm
meV}$ in the right lead implying the chemical potential difference $\Delta\mu =
0.1$~meV. Figure \ref{MBE_WOph}(b) shows the ME energy spectrum of the QD system, in
which the electron-electron interaction is included while no electron-photon coupling
has been introduced. Both the energies of SE states $N_e = 1$ (1ES, red dots) and
two-electron states $N_e = 2$ (2ES, blue dots) vary linearly proportional to the
applied plunger gate voltage $V_{\rm pg}$ but with different slopes.  The two-electron
states are located at relatively higher energies due to the Coulomb repulsion effect in
the QD-embedded system.

\begin{figure}[tbqh]
 \includegraphics[width=0.5\textwidth,angle=0]{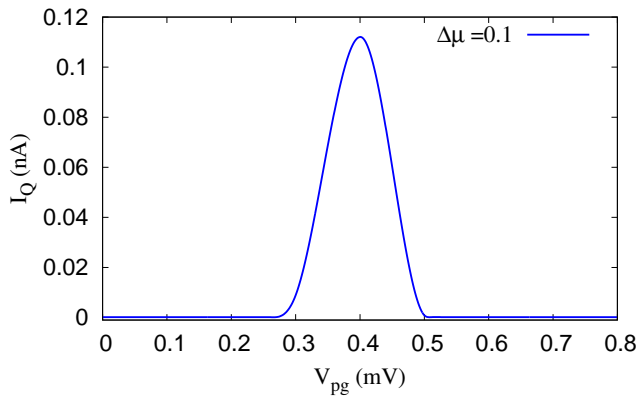}
\caption{(Color online) The net charging current $I_Q$ is plotted as a function of
plunger gate voltage $V_{\rm pg}$ at time $t = 220$~ps in the case of no photon cavity.
Other parameters are $B = 0.1~{\rm T}$ and $\Delta \mu =0.1~{\rm meV}$.
 } \label{I_Vpg_WOph}
\end{figure}

The SE state energy is tunable as a function of plunger gate voltage $V_{\rm pg}$
following $E_{\rm SE}(V_{\rm pg}) = E_{\rm SE}(0) + e V_{\rm pg}$.  
We rank the SE and ME states by energy. In the absence of
plunger gate voltage, the lowest active SE states in the central system are $|4)$ and
$|5)$ with energies $E_{\rm 4}(0) = 0.741$~meV and $E_{\rm 5}(0) = 0.744$~meV,
respectively. These two SE states may enter the chemical potential window
$[\mu_L,\mu_R] = [1.1, 1.2]$~meV by tuning the plunger gate voltage to be $V_{\rm pg}
\approx [0.35, 0.45]\ {\rm mV}$. Consequently, the SE states occupying the first
subband in the left lead are allowed to tunnel into the central ME system making
resonant tunneling from the left to the right lead manifesting a main-peak feature in
charging current $I_Q = 0.112$~nA at $V_{\rm pg}=0.4\ {\rm mV}$ as shown in
\fig{I_Vpg_WOph}.

\begin{figure}[htbq]
      \includegraphics[width=0.47\textwidth,angle=0]{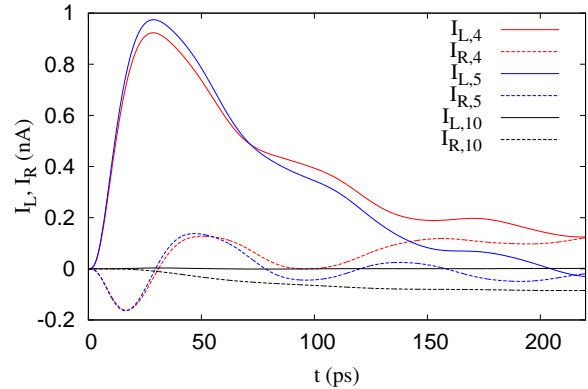}
      \caption{(Color online) Partial currents as a function of time without photon cavity:
      $I_L$ (red solid) and $I_R$ (red dashed) through the state $|4)$;
      $I_L$ (blue solid) and $I_R$ (blue dashed) through the state $|5)$. Other parameters
      are $V_{\rm pg} = 0.4$~mV, $\Delta \mu =0.1~{\rm meV}$, $B = 0.1~{\rm T}$,
      and $\hbar \Omega_0 = 2.0~{\rm meV}$.} \label{I_t_WOph}
\end{figure}

In \fig{I_t_WOph}, we show the time evolution of the left and right partial charging
currents in the case with no photon cavity to understand better how the $|4)$ and $|5)$
SE states in the bias window as well as the ground state 
with two electrons $|10)$ contribute to the transport. The state $|10)$ contributes  
because the energy difference $E_{10}-E_{4}$ (which includes the charging
energy) is also in the bias window. 
In the short-time regime at $t = 40$~ps, the partial current through the
three active ME states are $I_{L,4} = 0.852$~nA and $I_{R,4} = 0.1$~nA through the
state $|4)$ (red lines), $I_{L,5} = 0.906$~nA and $I_{R,5} = 0.121$~nA through the
state $|5)$ (blue lines), and $I_{L,10} = 0.002$~nA and $I_{R,10} = -0.025$~nA through
the state $|10)$ (black lines). As a result, the net partial current contributed by the
three active SE and ME states are $I_4 = 0.952$~nA, $I_5 = 1.027$~nA, $I_{10} = -0.023$~nA
resulting in $I_Q = 1.956$~nA. In the long-time regime at $t = 220$~ps, $I_{L,4} =
0.125$~nA and $I_{R,4} = 0.12$~nA through the state $|4)$ (red lines), $I_{L,5} =
-0.031$~nA and $I_{R,5} = -0.019$~nA through the state $|5)$ (blue lines), and
$I_{L,10} = 0.002$~nA and $I_{R,10} = -0.085$~nA through the state $|10)$ (black
lines).  The net partial charging current contributed by the three active ME states are
thus $I_4 = 0.245$~nA, $I_5 = -0.05$~nA, and $I_{10} = -0.083$~nA, thereby leading to
the net charging current $I_Q = 0.112$~nA. This exactly agrees with the result shown in
\fig{I_Vpg_WOph}. In the short-time regime, the left partial current contributed by the
states $|4)$ and $|5)$ is much large than the right partial current.  This illustrates
significant charge accumulation in the short-time regime and, hence, manifests a broad
peak structure as shown in \fig{I_t_WOph}

\begin{figure}[htbq]
 \includegraphics[width=0.23\textwidth]{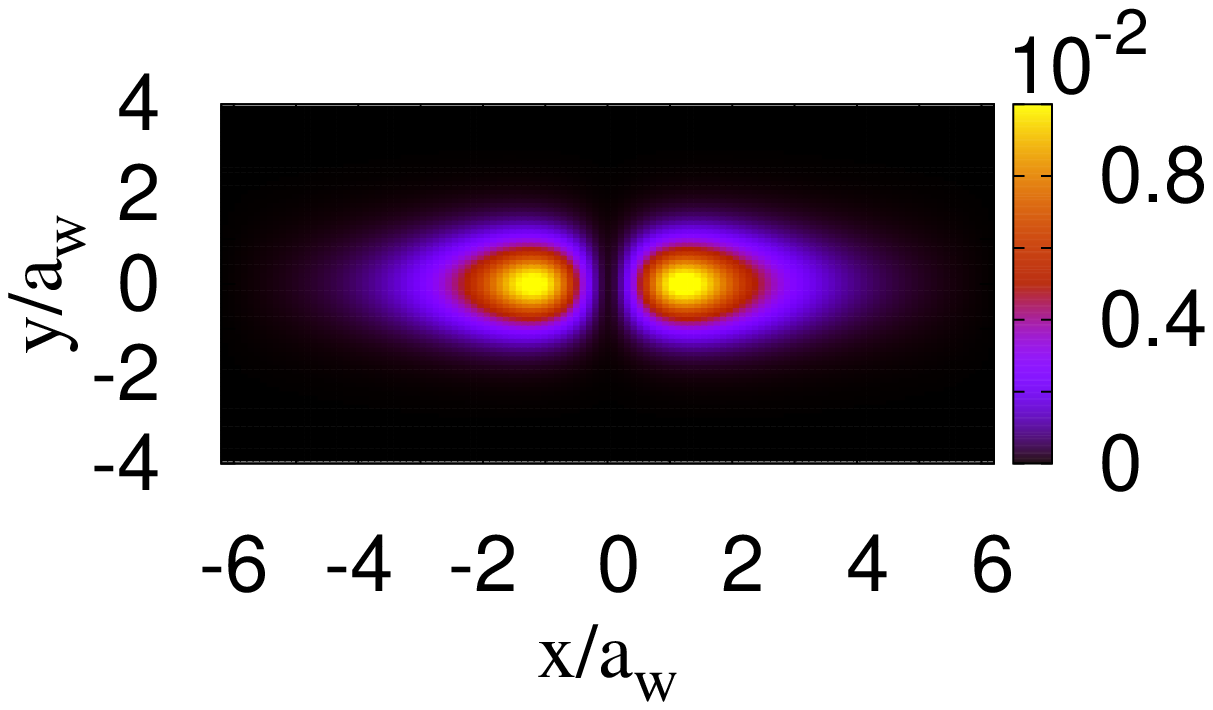}
 \includegraphics[width=0.23\textwidth]{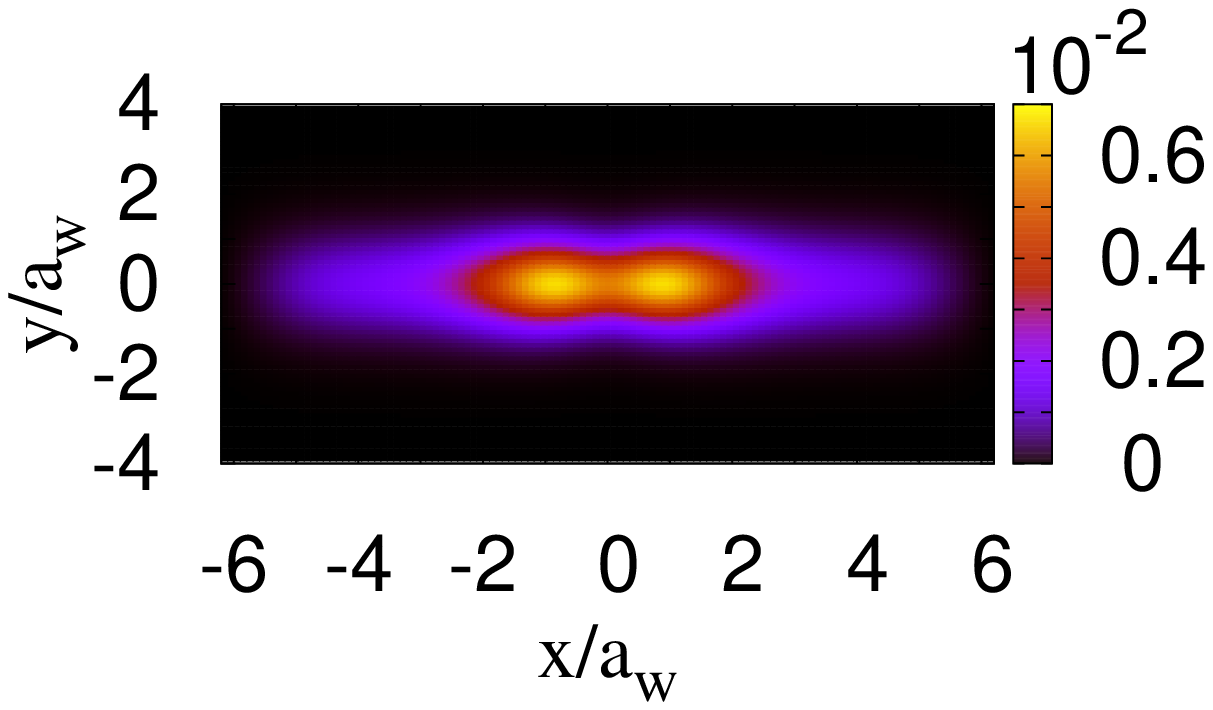}
 \caption{(Color online) The spatial distribution of the ME charge density at short-time
$t = 40$~ps (left panel) and long-time $t = 220$~ps (right panel) in the case with no
photon cavity, where the plunger gate is $V_{\rm pg} = 0.4$~meV. Other parameters are
$B=0.1$~T, $a_{w}$ = $23.8~{\rm nm}$, $L_x$ = $300$~nm = $12.6 a_w$, and $\Delta \mu
=0.1~{\rm meV}$.} \label{Q_WO_Ph}
\end{figure}

In order to understand the nature of the electrons traversing the QD-embedded system,
the distribution of ME charge is presented in \fig{Q_WO_Ph} in the short time regime
$40~{\rm ps}$ (left panel) and the long time regime $t = 220~{\rm ps}$ (right panel)
where the chemical potential difference is $\Delta \mu = 0.1$~meV.  In the short-time
regime, the electrons in the QD-embedded system exhibits longitudinal oscillations.
Two localized peaks are found located at
the edges in the transport direction of the embedded QD due to the breaking of the
translational invariance at the edges of the embedded QD, as shown in \fig{Q_WO_Ph}(a),
that favors the electrons making coherent elastic multiple scattering.  In the
long-time regime, a broader bound state with a long tail in the transport direction is
found that corresponds to the resonant state in the finite wire system. In the
following sections, we shall place the QD system in a photon cavity with a
single-photon mode. We shall analyze the transient transport properties for the cases
with linear polarizations in either $x$ or $y$ directions.

\subsection{$x$-polarized photon mode}

\begin{figure}[tbhq]
 \includegraphics[width=0.44\textwidth]{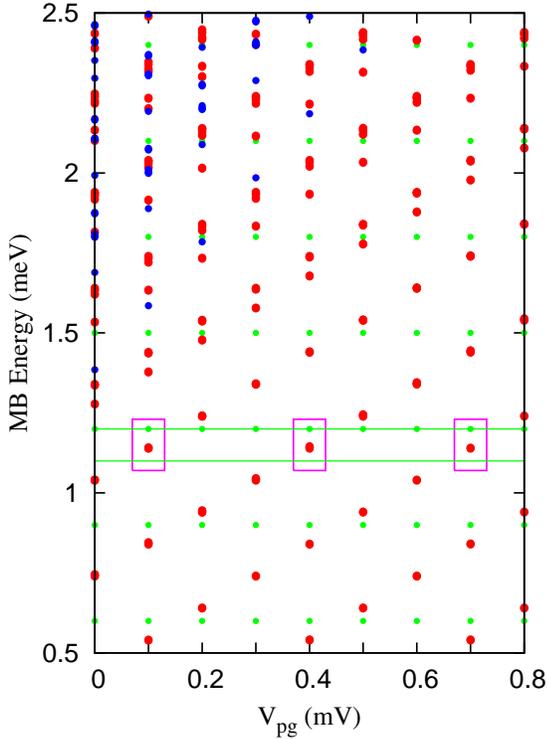}
 \caption{(Color online) MB Energy spectrum versus the plunger gate voltage $V_{\rm pg}$
 in the case of $x$-polarized photon field, where zero electron states ($N_e = 0$, green dots),
single-electron states ($N_e = 1$, red dots), and two-electron states ($N_e = 2$, blue
dots) are included.  Other parameters are $B=0.1$~T, $\Delta\mu = 0.1$~meV, and $\hbar
\omega_{\rm ph} = 0.3~{\rm meV}$. } \label{MBE_Wph_Xp}
\end{figure}

Here, we demonstrate how the QD embedded in a quantum wire can be controlled by the
plunger-gate and how it is influenced by the photon field, where the electric field
of the TE$_{011}$ mode is polarized in the $x$-direction.  The initial condition of the
system under investigation is an empty central system (no electron) that is coupled to
a single-photon mode with one photon present, connected to the leads with a source-drain bias.  
The MB energy
spectrum of the electron-photon interacting MB system is illustrated in
\fig{MBE_Wph_Xp}.  As shown in the previous section, active states get into the bias
window around $V_{\rm pg}^0 = 0.4$~mV in the case with no photon cavity.  It is
interesting to note that additional active states can be included around $eV_{\rm pg} =
eV_{\rm pg}^0 \pm \hbar \omega_{\rm ph}$ as is clearly seen in \fig{MBE_Wph_Xp}, this
implies that the $x$-polarized photon-field induced active propagating states can be
found around $V_{\rm pg} = 0.1$ and $0.7$~mV when the photon energy is $\hbar
\omega_{\rm ph} = 0.3$~meV.  The additional photon-induced propagating states play an
important role to enhance the electron tunneling from the leads to the QD system.

Figure \ref{I-g_{em}} shows the net charging current $I_Q$ as a function of the
plunger-gate voltage $V_{\rm pg}$ in the presence of the $x$-polarized photon field at
time $t = 220$~ps. We fix the photon energy at $\hbar\omega_{\rm ph} = 0.3~{\rm meV}$
and change the electron-photon coupling strength $g_{\rm ph}$.  A main peak around
$V_{\rm pg}^0 = 0.4$~mV is found, a robust left side peak around $eV_{\rm pg}= eV_{\rm
pg}^0 - \hbar\omega_{\rm ph}$ is clearly shown, and a right side peak around $eV_{\rm
pg}= eV_{\rm pg}^0 + \hbar\omega_{\rm ph}$ can be barely recognized.  The left side
peak exhibits photon-assisted transport feature from the SE MB states $|\breve{20})$
and $|\breve{22})$ in the bias window by absorbing a photon energy $\hbar\omega_{\rm
ph}$ to the SE MB states $|\breve{26})$ and $|\breve{28})$ above the bias window.
However, the opposite photon-assisted transport feature caused by a photon emission (the
right side peak) is significantly suppressed.
\begin{figure}[tbhq]
 \includegraphics[width=0.5\textwidth]{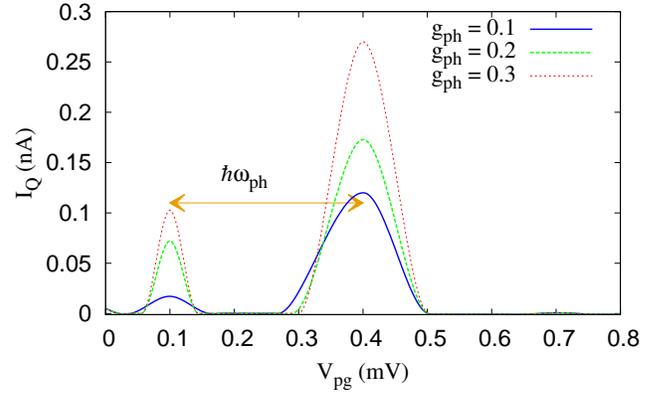}
\caption{(Color online) The net charging current $I_Q$ versus the plunger gate voltage
 $V_{\rm pg}$ in the case of $x$-polarized photon field at time $t = 220~{\rm ps}$
 with different electron-photon coupling strength:
 $g_{\rm ph} = 0.1$~meV (blue solid), $g_{\rm ph} = 0.2$~meV (green dashed),
 and $g_{\rm ph} = 0.3$~meV (red dotted).
 Other parameters are $\hbar\omega_{ph} = 0.3~{\rm meV}$, $\Delta \mu =0.1~{\rm meV}$, and $B = 0.1~{\rm T}$.
 }
 \label{I-g_{em}}
\end{figure}

The main charge current peaks for $V_{\rm pg} = 0.4$~mV are $I_Q^M = 0.120, 0.173$, and
$0.270$~nA corresponding to $g_{\rm ph} = 0.1~{\rm meV}$, (blue solid), $g_{\rm ph} =
0.2~{\rm meV}$ (green dashed), and $g_{\rm ph} = 0.3~{\rm meV}$ (red dotted) as shown
in \fig{I-g_{em}}. Our results demonstrate that the current carried by the 
electrons with energy within the
bias window can be strongly enhanced by increasing the electron-photon coupling
strength. At $V_{\rm pg} = 0.1$~mV, the left side peaks in the charging current are
$I_Q^S = 0.017, 0.072$, and $0.103$~nA corresponding to $g_{\rm ph} = 0.1, 0.2$, and
$0.3$~meV. This implies that the electrons may absorb a single-photon energy and,
hence, the charging current manifests a photon-assisted transport.

\begin{figure}[htbq]
      \includegraphics[width=0.47\textwidth]{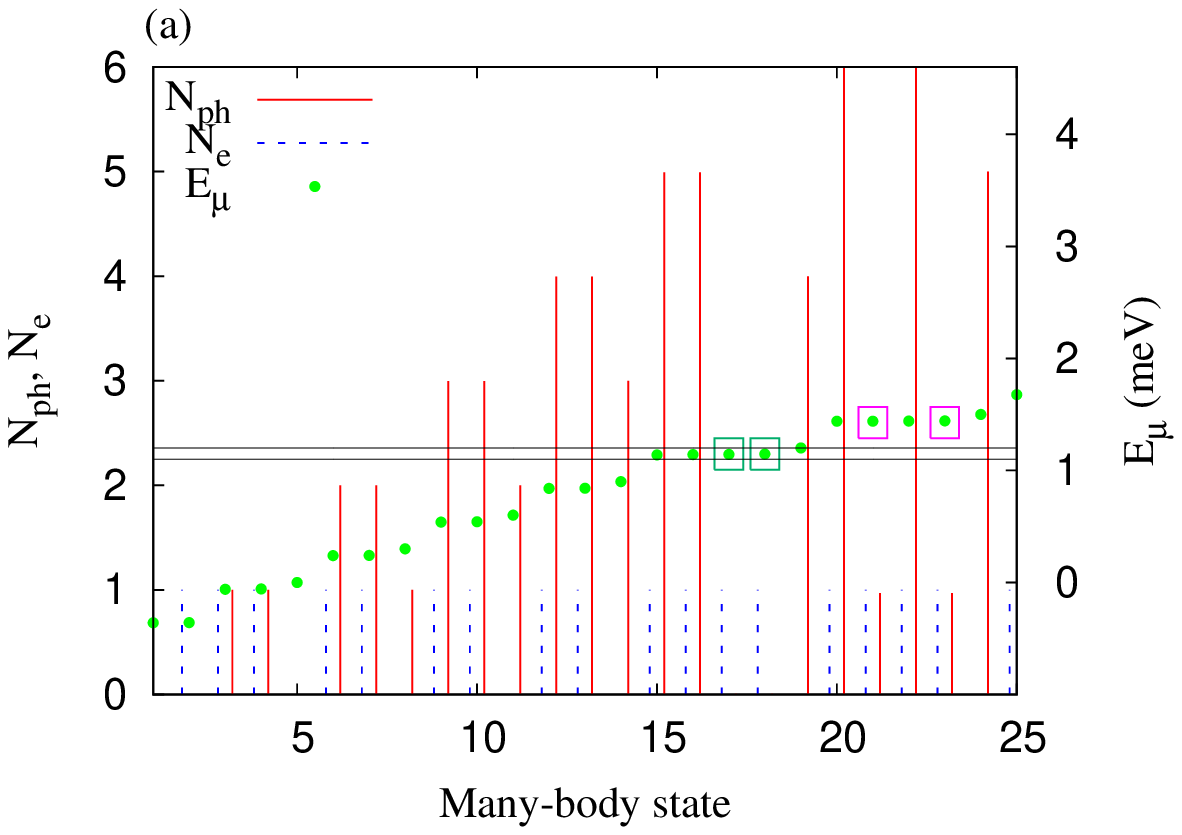}
      \includegraphics[width=0.47\textwidth]{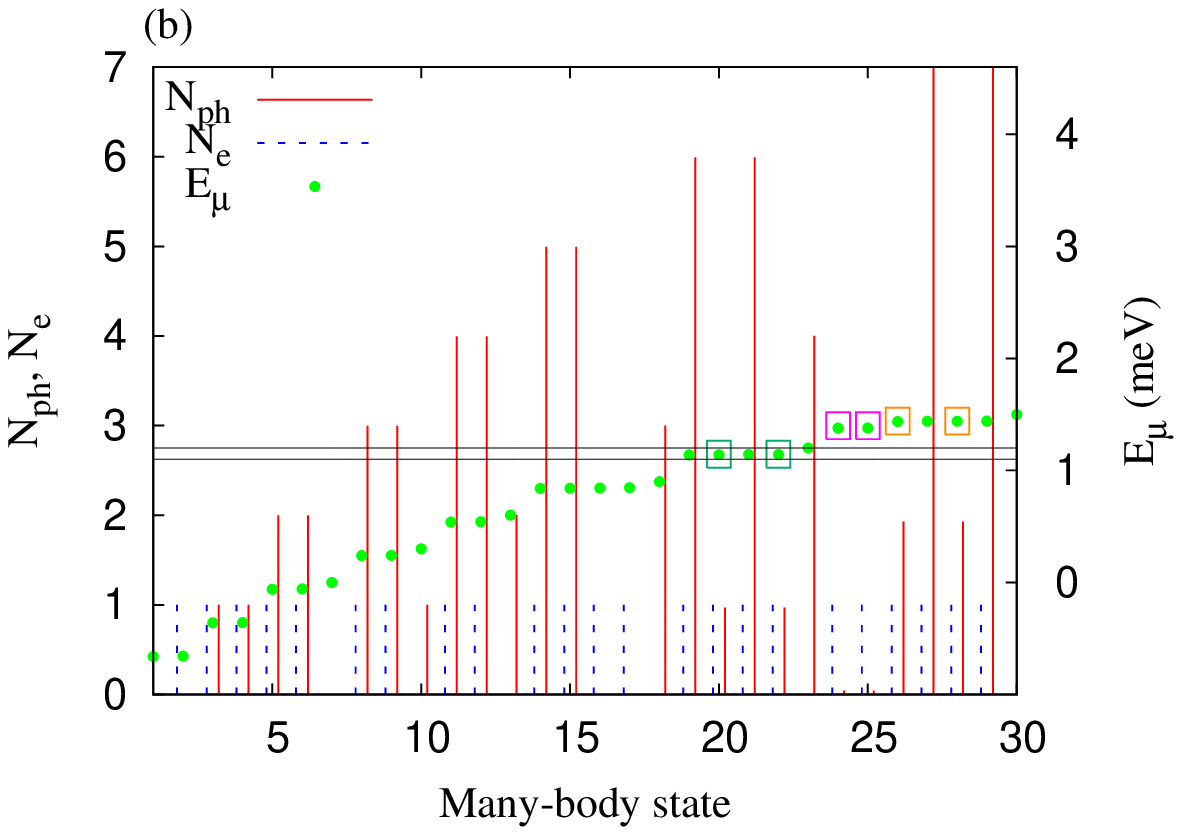}
      \caption{(Color online) The MB energy spectrum $E_{\mu}$ (dotted green), the mean
               electron number in the MB state $|\breve{\mu})$ (blue dashed line),
               the mean photon number $N_{\rm ph}$ (red line) in the case of
               $x$-polarized field:
               (a) $V_{\rm pg} = 0.4$~meV and (b) $V_{\rm pg} = 0.7$~meV.
               Other parameters are $B = 0.1~{\rm T}$, $\Delta \mu =0.1~{\rm meV}$,
               $g_{\rm ph} = 0.1$~meV, $\hbar \omega_{\rm ph} = 0.3$~meV.
               }
\label{NeNphSz_Vpg04}
\end{figure}

To identify the active MB states contributing to the transient transport, we show the
characteristics of the MB states at $V_{\rm pg} = 0.4$ and 0.1~meV in
\fig{NeNphSz_Vpg04}(a) and (b) corresponding, respectively, to the main peak and the
left side peak in $I_Q$ shown in \fig{I-g_{em}}. More precisely, there are five MB
states contributing to the main peak in $I_Q$ at $V_{\rm pg} = 0.4$~meV.  The five
active MB states are: $|\breve{17})$ and $|\breve{18})$ with energies $E_{17} =
1.143$~meV and $E_{18} = 1.145$~meV in the bias window ($N_e = 1$, $N_{\rm ph} =
0.04$), $|\breve{21})$, $|\breve{23})$ with energies $E_{21} = 1.439$~meV and $E_{23} =
1.441$~meV above the bias window ($N_e = 1$, $N_{\rm ph} = 0.96$) shown in
\fig{NeNphSz_Vpg04}(a), and $|\breve{53})$ with energy $2.488$~meV (not shown). It is
interesting to notice that $E_{17} + \hbar\omega_{\rm ph} \cong E_{21}$ and $E_{18} +
\hbar\omega_{\rm ph} \cong E_{23}$, this implies a photon-assisted transport through
the higher MB states.

When an electron enters the QD system it interacts with the photon in the cavity. Its
energy is thus not in resonance with the electron states in the bias window, but with
the electron states, photon replicas, which are a with photon energy $\hbar
\omega_{ph}$ above the states in the bias window. The photon activated states above the
bias contain approximately one more photon than the states in the bias window and,
hence, the main-peak in $I_Q$ is mainly due to a single-photon absorption mechanism. In
addition to the main-peak feature at plunger-gate voltage $V^{\rm M}_{\rm pg}$, two
side peaks can be recognized at $eV^{\rm S}_{\rm pg} = eV^{\rm M}_{\rm pg} \pm
\hbar\omega_{\rm ph}$ induced by a photon-assisted transport, where the system
satisfies $e\Delta V^{\rm MS}_{\rm pg}  = e|V^{\rm M}_{\rm pg}-V^{\rm S}_{\rm pg}|
\cong \hbar\omega_{\rm ph} $. It has been pointed out that this plunger-gate controlled
photon-assisted transport is repeatable with period related to the Coulomb charging
energy.\cite{Kouwenhoven73.3443}

Figure \ref{NeNphSz_Vpg04}(b) shows how the left-side peak in the net charging
current $I_Q$ shown previously in \fig{I-g_{em}} is contributed by the MB states.
First, the left current $I_{L} = 0.001$~nA and the right current $I_R = -0.001$~nA
contributed by the $|\breve{20})$ and $|\breve{22})$ MB states (green squared dot)
containing $N_{\rm e} = 1$ and $N_{\rm ph} = 0.96$ within the bias window are almost
negligible, this implies the left side peak in $I_Q$ is not induced by the resonant
tunneling effect. Second, the $|\breve{24})$ and $|\breve{25})$ MB states (pink squared
dot) contain $N_{\rm e} = 1$ and $N_{\rm ph} = 0.04$ with energies $E_{24} = 1.376$~meV
and $E_{25} = 1.379$~meV, above the bias window. These two states contribute,
respectively, to the charging current $I_{24} = 0.0$~nA ($I_{L,24} = 0.003$~nA,
$I_{R,24} = -0.003$~nA) and $I_{25} = 0.001$~nA ($I_{L,25} = 0.007$~nA, $I_{R,25} =
-0.006$~nA) and, hence, generate a charging current $I_Q^c =
0.001$~nA. Third, the $|\breve{26})$ and $|\breve{28})$ MB states (orange squared dot)
contain $N_{\rm e} = 1$ and $N_{\rm ph} = 1.96$ with energies $E_{26} = 1.435$~meV and
$E_{28} = 1.438$~meV above the bias window. These two states contribute, respectively,
to the charging current $I_{26} = 0.01$~nA ($I_{L,26} = 0.010$~nA, $I_{R,26} = 0.0$~nA)
and $I_{28} = 0.004$~nA ($I_{L,28} = 0.005$~nA, $I_{R,28} = -0.001$~nA) and, hence
generate a photon-assisted tunneling current $I_Q^{\rm ph} = 0.014$~nA.  The main
contribution of the left side peak in the charging current is then $I_Q \approx I_Q^c +
I_Q^{\rm ph} = 0.015$~nA, this coincides with the result shown in \fig{I-g_{em}}.

\begin{figure}[htbq]
 \includegraphics[width=0.22\textwidth,angle=0]{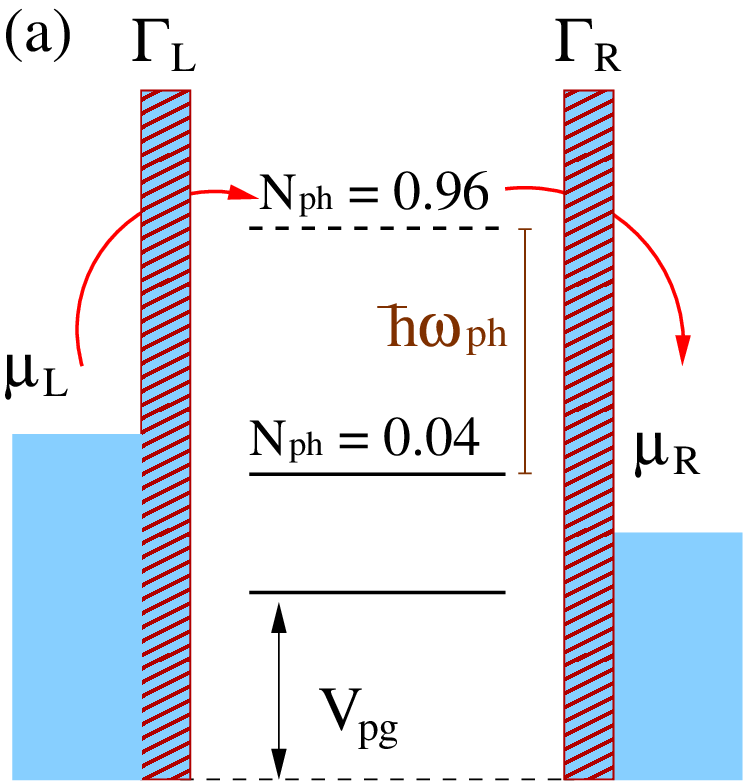}
 \includegraphics[width=0.22\textwidth,angle=0]{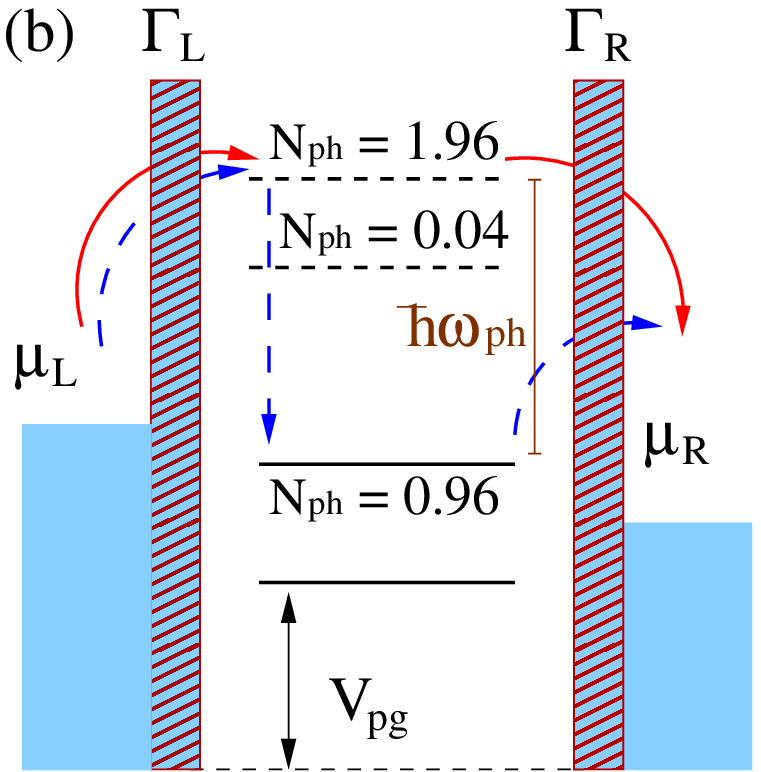}
\caption{(Color online) Schematic representation of photon activated resonance energy
levels and electron transition by changing the plunger gate voltage $V_{\rm pg}$ at the
main peak (a) and the left side peak (b) in \fig{I-g_{em}}.  The QD system is embedded
in a photo cavity with the photon energy $\hbar \omega_{\rm ph}$ and photon content
$N_{\rm ph}$ in each many-body state.  The chemical potential difference is $\Delta \mu
= \mu_L - \mu_R$, and $\Gamma_{\rm {L,R}}$ is the coupling strength between the QD
system and the leads.} \label{E_Level}
\end{figure}

The schematic diagram in \fig{E_Level} is shown to illustrate the dynamical
photon-assisted transport processes involved in the formation of the main peak and the
left side peak in the net charging current $I_Q$ shown in \fig{I-g_{em}}.  It is
illustrated in \fig{E_Level}(a) that the transport mechanism forming the main peak in
$I_Q$ is mainly due to the photon-assisted tunneling to the MB states above the 
bias window containing approximately a single photon.  Figure \ref{E_Level}(b)
represents two main transport mechanisms forming the left side peak in $I_Q$.  The
electrons in the left lead may absorb two photons to the MB states containing
approximately two photons above the bias window. After that, the electrons may either
perform resonant tunneling to the right lead (red solid arrow) or make multiple
inelastic scattering by absorbing and emitting photon energy $\hbar\omega_{\rm ph}$ in
the QD system (blue dashed arrow). This is the key result of this paper.

\begin{figure}[htbq]
  \includegraphics[width=0.23\textwidth]{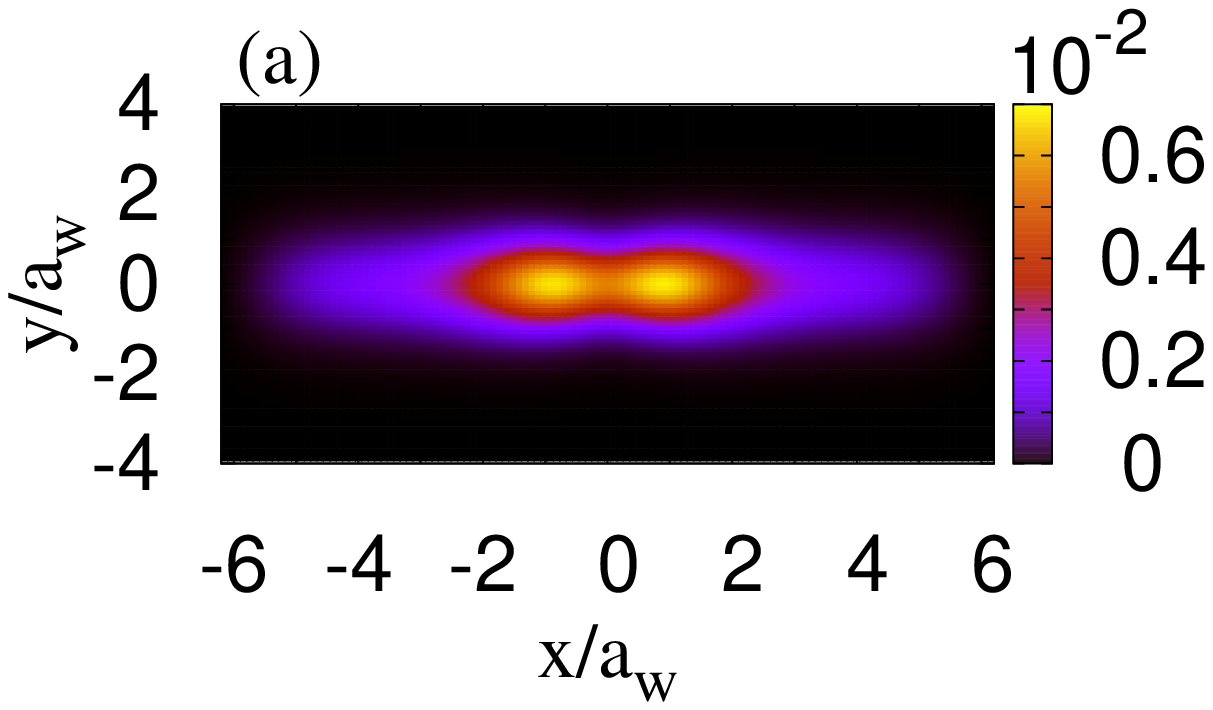}
 \includegraphics[width=0.23\textwidth]{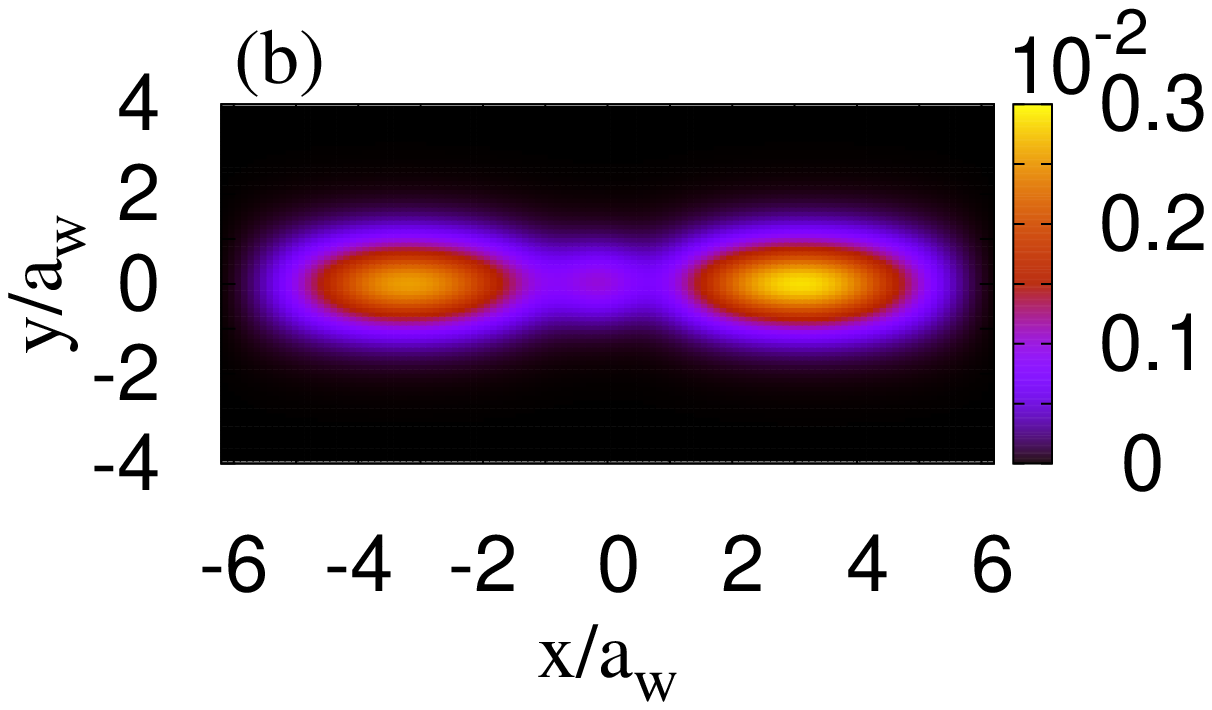}
\caption{(Color online) The spatial distribution of the many-electron charge density of
the QD system with $x$-polarized photon field at time $220$~ps corresponding to the
main peak (a) and the left side peak (b) for the case of $g_{\rm ph} = 0.1$~meV shown
in the \fig{I-g_{em}} (blue solid line). Other parameters are $\hbar \omega_{\rm ph} =
0.3~{\rm meV}$, $B=0.1$~T, $a_{w} = 23.8~{\rm nm}$, $L_x = 300~{\rm nm}$, and $\hbar
\Omega_0 = 2.0~{\rm meV}$.} \label{Q_W_Ph}
\end{figure}

To get better insight into the dynamical electronic transport, the spatial distribution
of the ME charge at $t=220$~ps is shown in \fig{Q_W_Ph}. Similar to the QD system in
the absence of the photon cavity, the ME charge distribution at the main-peak in $I_Q$
forms resonant peaks at the edges of the QD, as shown \fig{Q_W_Ph}(a), that is related
to an antisymmetric state in the QD. The partial occupation contributed by the photon
activated resonant MB states $|\breve{21})$ and $|\breve{23})$ are $0.432e$ and
$0.454e$, respectively. Comparing to the case with no photon cavity, the slight
enhancement in the ME charge indicates that the tunneling of electrons into the QD
system becomes faster in the presence of the photon cavity and, hence, the charging
current is enhanced. It is shown in \fig{Q_W_Ph}(b) that the ME charge in the case of
side peak in $I_Q$ manifests an extended SE state, which is formed outside the QD. The
partial occupation contributed by the photon activated resonant MB states
$|\breve{24})$ and $|\breve{25})$ are $0.018e$ and $0.025e$, respectively. By
increasing the photon energy $\hbar\omega_{\rm ph}$, the left side peak in $I_Q$ can be
enhanced and is shifted to lower energy (not shown). The slight asymmetry seen in
the charge distribution in \fig{Q_W_Ph}(b) is caused by the $x$-polarized electric
field of the photons. 

\subsection{$y$-polarized photon mode}

We consider here the TE$_{101}$ $y$-polarized photon mode, where the electric field
of the photons is
perpendicular to the transport direction through the QD system.  The QD system is
assumed to be initially containing no electron $N_e = 0$, but one photon in the cavity 
$N_{\rm ph} =1$. Since our system is considered to be anisotropic, elongated in the
$x$-direction, we shall demonstrate that the photon-assisted transport effect is 
much weaker in the case of a $y$-polarized photon mode in comparison with that of
$x$-polarization discussed in the previous section.

\begin{figure}[htbq]
 \includegraphics[width=0.44\textwidth]{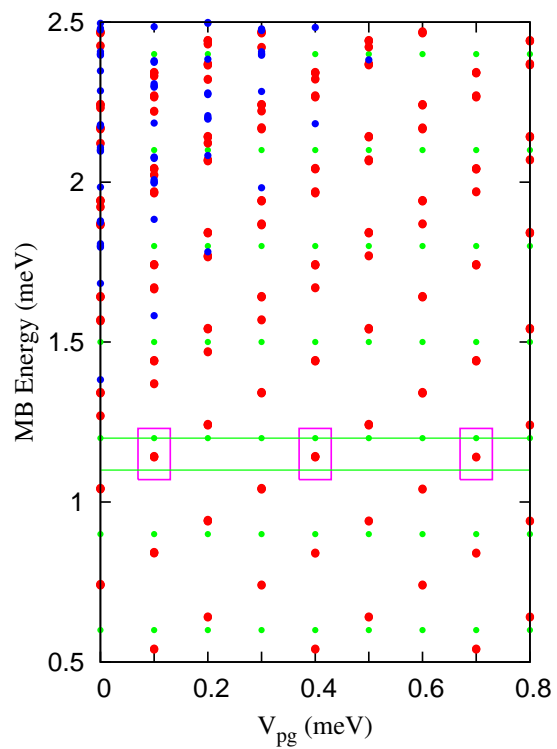}
\caption{(Color online) MB energy spectrum versus the plunger-gate voltage $V_{\rm pg}$
in the case of $y$-polarized photon field: zero-electron states $N_e = 0$ (green dots),
single-electron states $N_e = 1$ (red dots),  and two-electron states $N_e = 2$ (blue
dots). Other parameters are $B=0.1$~T, $\Delta\mu = 0.1$~meV, $\hbar \Omega_{0} =
2.0$~meV, $\hbar \omega_{\rm ph} = 0.3~{\rm meV}$, and $g_{\rm ph} = 0.1$.}
 \label{MBE_Wph_Yp}
\end{figure}

In \fig{MBE_Wph_Yp}, we present the MB energy spectrum as a function of plunger-gate
voltage $V_{\rm pg}$ for a QD system influenced by the $y$-polarized field with photon
energy $\hbar \omega_{\rm ph} = 0.3~{\rm meV}$. Besides the propagating state at
$V_{\rm pg}= 0.4$~mV within the bias window (green lines), there are two additional
electronic propagating states appearing at $V_{\rm pg}= 0.1$ and 0.7~mV caused by the
presence of the photon field as marked by the squared dots shown in the figure.

\begin{figure}[htbq]
 \includegraphics[width=0.49\textwidth]{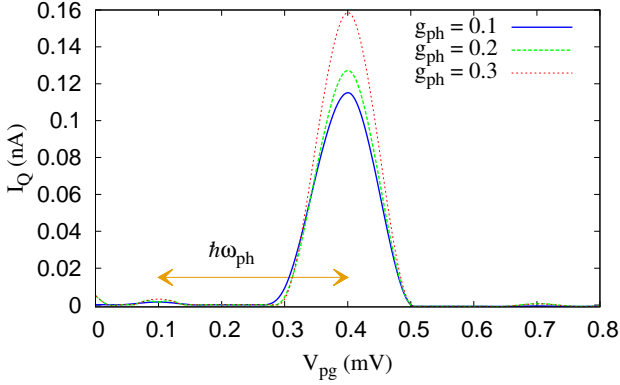}
\caption{(Color online)  net charging current versus the plunger gate voltage $V_{\rm
pg}$ at time ($t = 220~{\rm ps}$) in the case of $y$-polarized photon field.  The
electron-photon coupling is changed to be $g_{\rm ph} = 0.1$~meV (blue solid), $g_{\rm
ph} = 0.2$~meV (green dashed), and $g_{\rm ph} = 0.3$~meV  (red dotted).
 Other parameters are $\hbar\omega_{\rm ph} = 0.3~{\rm meV}$, $\Delta \mu =0.1~{\rm meV}$,
 and $B = 0.1~{\rm T}$.
 }
\label{I-Yp}
\end{figure}

Figure \ref{I-Yp} shows the net charging current in the case of $y$-polarized photon
field, in which there is initially one photon $N_{\rm ph} = 1$ with energy
$\hbar\omega_{\rm ph} = 0.3~{\rm meV}$ fixed while the electron-photon coupling
strength is changed. It is seen that main peak currents at $V_{\rm pg} = 0.4$~mV are:
$I_Q^M = 0.115$~nA for $g_{\rm ph} = 0.1$~meV (blue solid), $I_Q^M = 0.127$~nA for
$g_{\rm ph} = 0.2$~meV (green dashed), and $I_Q^M = 0.159$~nA for $g_{\rm ph} =
0.3$~meV (red dotted). Moreover, weak left side-peak current at $V_{\rm pg} = 0.1$~mV
can be recognized: $I_Q^S = 1.0$~pA for $g_{\rm ph} = 0.1$~meV, $I_Q^S = 1.7$~pA for
$g_{\rm ph} = 0.2$~meV, and $I_Q^S = 3.2$~pA for $g_{\rm ph} = 0.3$~meV. We notice that
both the side and main peak currents are enhanced when the electron-photon coupling
strength is increased.  In order to get better understanding of the current
enhancement, we repeat the analysis of the photon activated MB energy states
contributing to the electronic transport.

\begin{figure}[htbq]
      \includegraphics[width=0.45\textwidth]{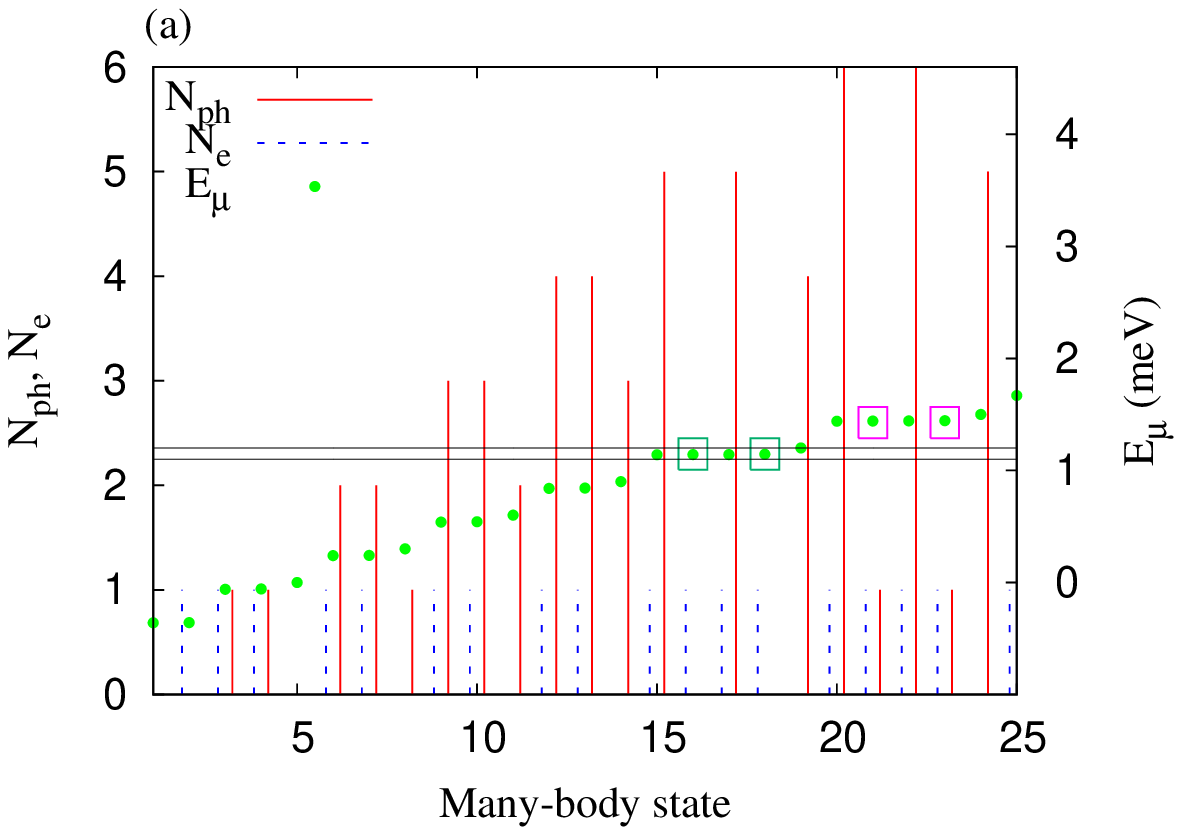}
      \includegraphics[width=0.45\textwidth]{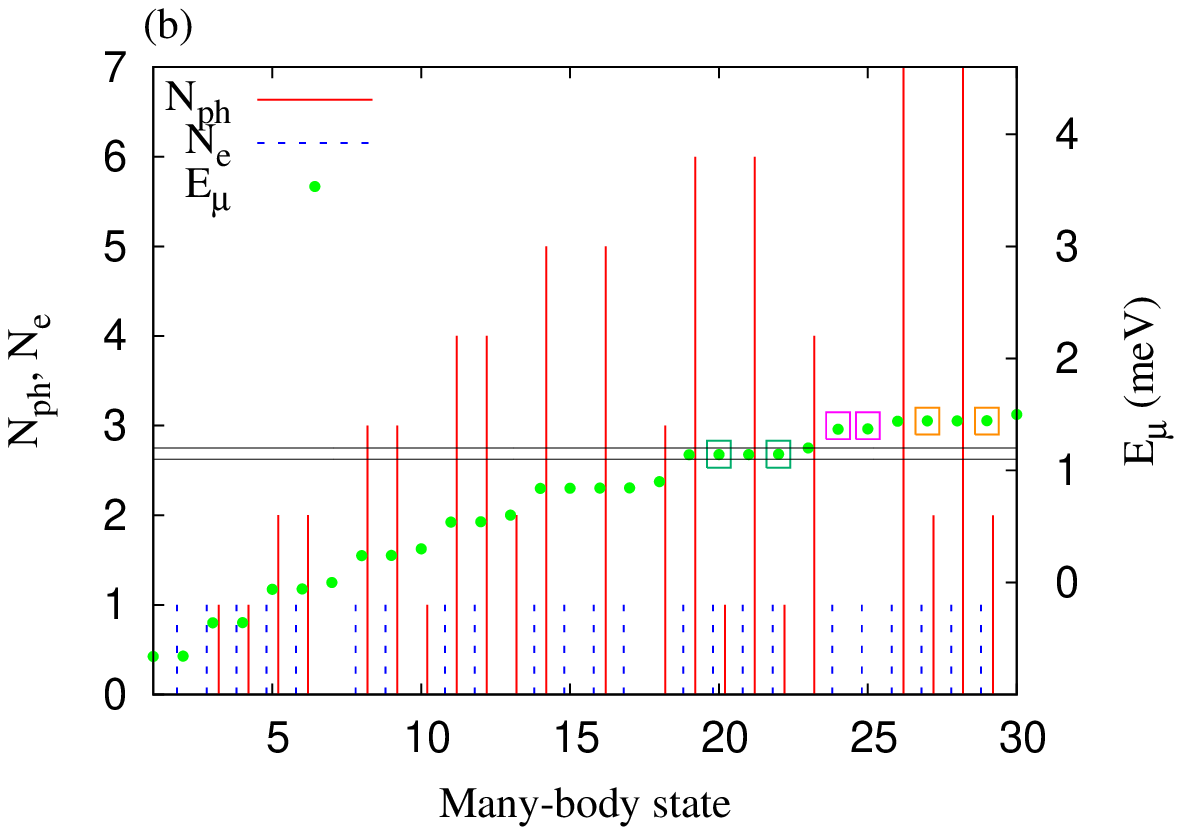}
      \caption{(Color online) The many-body energy spectrum $E_{\mu}$ (dotted green), the mean
               electron number in the many-body state $|\breve{\mu})$ (blue dashed line),
               the mean photon number $N_{ph}$ (red line)
               of the main peak $V_{\rm pg} = 0.4$~meV (a),
               and the left side peak $V_{\rm pg} = 0.1$~meV (b).
               The magnetic fields $B = 0.1~{\rm T}$, $\Delta \mu =0.1~{\rm meV}$,
               $g_{\rm ph} = 0.1$~meV, $\hbar \omega_{\rm ph} = 0.3$~meV.
               In the case of $y$-polarized photon field.}
\label{NeNphSz_Vpg04_Yp}
\end{figure}

In \fig{NeNphSz_Vpg04_Yp}(a), we show the MB states at $V_{\rm pg} = 0.4$~mV and
$g_{\rm ph} = 0.1$. The active MB states are $|\breve{16})$ and $|\breve{18})$ with
energies 1.141 and 1.144~meV in the bias window ($N_{\rm ph} = 0$), $|\breve{21})$ and
$|\breve{23})$ with energies 1.441 and 1.444~meV above the bias window ($N_{\rm ph} =
1$), and $|\breve{53})$ with energy 2.483~meV ($N_{\rm ph} = 1$). It should be noticed
that $E_{16} + \hbar\omega_{\rm ph} \cong E_{21}$ and $E_{18} + \hbar\omega_{\rm ph}
\cong E_{23}$ indicating that these two MB states above the bias window are
photon-activated states. Furthermore, the higher active MB state with energy
approximately the same with the characteristic Coulomb energy, that is $E_{53}\approx
E_{\rm C}$, indicates a correlation induced active two-electron state.

The net charging current at $V_{\rm pg} = 0.4$~mV exhibiting the main current peak in
\fig{I-Yp} at $t = 220$~ps is mainly contributed by the MB states $|\breve{21})$
($I_{L,21} = 0.127$~nA, $I_{R,21} = 0.125$~nA) and $|\breve{23})$ ($I_{L,23} =
-0.032$~nA, $I_{R,23} = -0.018$~nA).  This indicates that the electrons in the left
lead can absorb one photon to the state $|\breve{21})$ and then emit one photon
preforming resonant tunneling to the right lead, and contribute to the charging current
$I_{21} = 0.252$~nA. Moreover, an opposite transport mechanism can happen for the
electrons in the right lead through the state $|\breve{23})$, and then contribute to
the charging current $I_{23} = -0.05$~nA. The scattering processes through these two
states results in a photon-assisted tunneling current $I_Q^{\rm ph} = 0.202$~nA. A
small current through $|\breve{53})$ is found due to the charging effect, namely $I_L =
0.002$~nA and $I_R = -0.087$~nA, and hence contribute to the charging current $I_Q^c =
-0.085$~nA due to charging effect. The contribution to the main peak in charging
current is therefore $I_Q \approx I_Q^{\rm ph} + I_Q^c$ = $0.117$~nA. This analysis is
consistent with the result shown in \fig{I-Yp}.

In \fig{NeNphSz_Vpg04_Yp}(b), we show the MB states at $V_{\rm pg} = 0.1$~mV and
$g_{\rm ph} = 0.1$.  The active MB states are: $|\breve{20})$ and $|\breve{22})$ with
energies $E_{20} = 1.141$~meV and $E_{22} = 1.144$~meV in the bias window ($N_{\rm ph}
= 1$); $|\breve{24})$ and $|\breve{25})$ with energies 1.368 and 1.371~meV above the
bias window ($N_{\rm ph} = 0$); and $|\breve{27})$ and $|\breve{29})$ with energy
$E_{27} = 1.441$~meV and $E_{29} = 1.444$~meV ($N_{\rm ph} = 2$). We notice that
$E_{20} + \hbar\omega_{\rm ph} \cong E_{27}$ and $E_{22} + \hbar\omega_{\rm ph} \cong
E_{29}$. This implies that the two MB states $|\breve{27})$ and $|\breve{29})$ above
the bias window are photon-activated states.

In \fig{I-Yp}, the net charging current at $V_{\rm pg} = 0.1$~mV manifests a small
side-peak current $I_Q^S = 1.0$~pA at $t = 220$~ps. This left side-peak structure in
$I_Q$ is mainly contributed by the MB states $|\breve{20})$ ($I_L = 1.1$~pA, $I_R =
-0.9$~pA) and $|\breve{22})$ ($I_L = 1.2$~pA, $I_R = -0.9$~pA) in the bias window.
These two states contribute to the resonant tunneling current, $I_Q^r = 0.5$~pA, that
is related to the charge accumulation effect. In addition, the states $|\breve{27})$
($I_{27} = 4 \times 10^{-3}$~pA) and $|\breve{29})$ ($I_{29} = 2\times 10^{-3}$~pA)
contribute to very weak charging current $I_Q^{\rm ph} = 6\times 10^{-3}$~pA due to
photon-assisted tunneling. The contribution to the side-peak current is therefore
$I_Q^S \approx I_Q^r + I_Q^{\rm ph}$ = $0.51$~pA. The suppression of the side-peak
current in the case of $y$-polarization is due to the anisotropy of our system.
The dipole momentum in the $y$-direction is much smaller in the $x$-direction,
and so is the electron-photon interaction strength.

\begin{figure}[htbq]
  \includegraphics[width=0.23\textwidth]{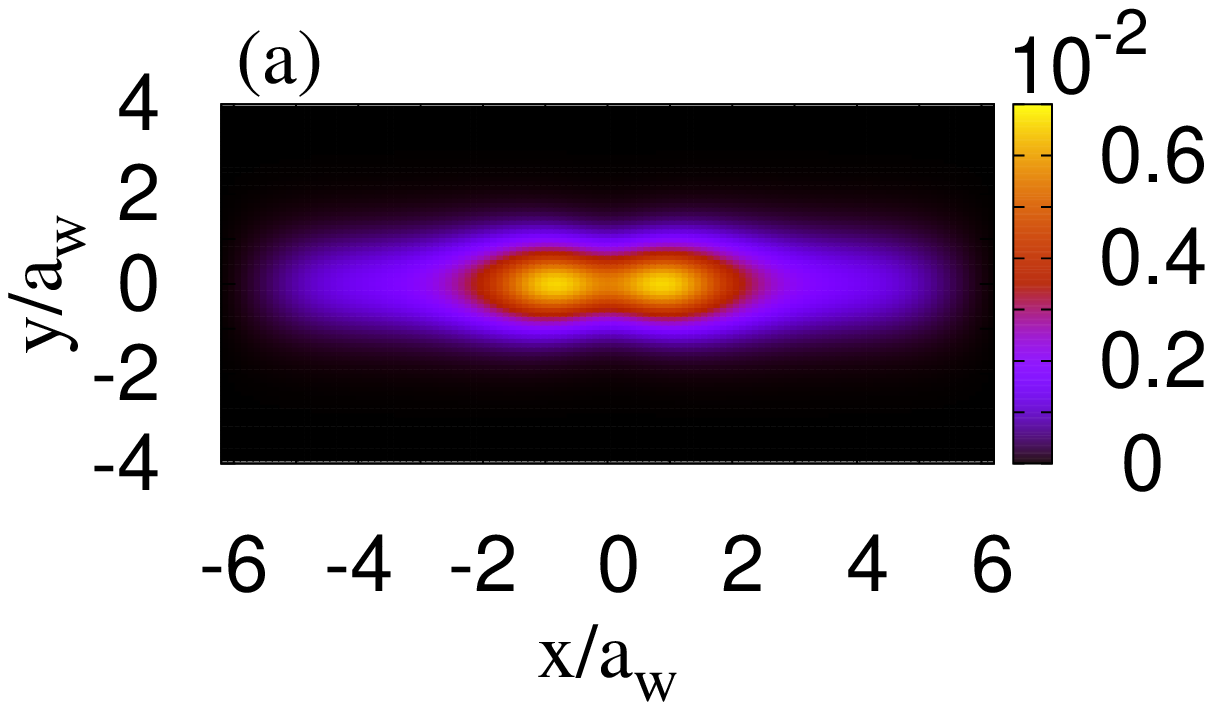}
  \includegraphics[width=0.23\textwidth]{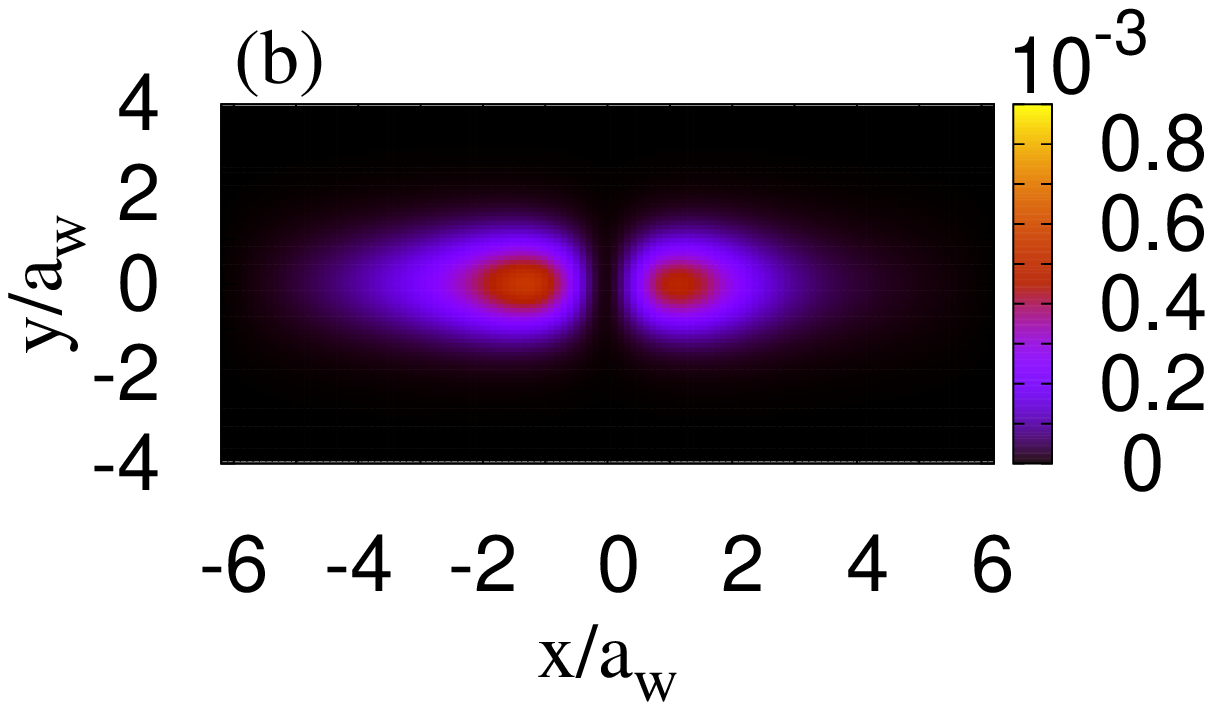}
\caption{(Color online) The spatial distribution of the ME charge density in the case
of $y$-polarized photon field at time $220$~ps: (a) $V_{\rm pg} = 0.4$~mV and (b)
$V_{\rm pg} = 0.1$~mV corresponding, respectively, to the main peak and the left side
peak in \fig{I-Yp} (blue line, $g_{\rm ph}=0.1$~meV). Other parameters are $\hbar
\omega_{\rm ph} = 0.3~{\rm meV}$, $B=0.1$~T, $a_{w} = 23.8~{\rm nm}$, $L_x = 300~{\rm
nm}$, and $\hbar \Omega_0 = 2.0~{\rm meV}$.}
 \label{Q_W_Ph_Y}
\end{figure}
The ME charge distribution in the presence of the $y$-polarized photon mode is shown in
\fig{Q_W_Ph_Y}.  It is seen that the main-peak current in \fig{I-Yp} forms an
elongated broad bound state in the central system due to the electron-photon
interaction as shown in \fig{Q_W_Ph_Y}(a). Moreover, the side-peak current in
\fig{I-Yp} forms a photon-assisted resonant state at the edges of the QD embedded in
the quantum wire, as is shown in \fig{Q_W_Ph_Y}(b).  We notice that the charge
distribution maxima around $x \approx \pm a_w$ of the main peak in $I_Q$ at $V_{\rm pg}
= 0.4$~mV with $g_{\rm ph}=0.1$~meV are almost the same in the cases without and with
photon mode.  As a consequence, the main-peak current $I_Q^M \simeq 0.1$~nA for theses
cases. Furthermore, the charging current maxima are located around $x \approx \pm 3
a_w$ in the case of $x$-polarization while located around $x \approx \pm 2 a_w$ in the
case of $y$-polarization.  The charge distribution maxima in the case of
$x$-polarization is closer to the edges of the central system implying the higher left
side-peak current at $V_{\rm pg} = 0.1$~mV.

\section{Concluding Remarks}\label{Sec:IV}

We have performed numerical calculation to investigate the transient current and charge
distribution of electrons through a QD embedded in a finite wire coupled to a
single-photon mode with $x$- or $y$-polarization.  A non-Markovian theory is utilized
where we solve a generalized QME that includes the electron-electron Coulomb interaction
and electron-photon coupling.  Initially, we examine the case without a photon cavity.
In the short-time regime, the charging current exhibits significant charge accumulation
effect. In the long-time regime, the charging current is suppressed due to the Coulomb
blocking effect. Furthermore, we have analyzed the photon-assisted current and the
characteristics of photon activated MB states with various parameters coupled to
single-photon mode in the photon cavity.  The photon-assisted current peaks
are enhanced by increased electron-photon coupling strength.

In the case of a QD system coupled to an $x$-polarized photon mode, the main current peak
is enhanced by the electron-photon coupling. The electrons may
absorb a single photon manifesting a photon-assisted secondary peak
which also incorporates correlation effects. In the case of a QD coupled to a $y$-polarized
photon mode, the main current peak is contributed to by two photon-activated
single-electron states and a correlation-induced two-electron state. The secondary peak
current in the case of $y$-polarization is suppressed due to the anisotropy of our
system.

The cavity photon assisted or enhanced transport here was attainable by selecting a
narrow bias window in order to facilitate the resonant placement and isolation of spin-pair 
of states with a single-electron component by the plunger gate in the bias window.
The bias window was kept in the lowest part of the MB energy spectrum and the low 
photon energy guarantees in most cases that only states close to this very descrete 
part of the spectrum are relevant to the transport. This is in contrast to our experience
with large bias window where the coupling to the cavitiy photons most often reduce the 
charging of the central system.\cite{Vidar85.075306,Vidar61.305,Arnold13:035314}

Our proposed plunger-gate controlled transient current in a single-photon-mode
influenced QD system should be observable due to recent rapid progress of measurement
technology.\cite{Feve1169.316}  The realization of a single-photon influenced QD device and
the generation of plunger-gate controlled transient transport may be useful in quantum
computation applications.

\ \\
\begin{acknowledgments}
This work was financially supported by the Icelandic Research and Instruments Funds,
the Research Fund of the University of Iceland, and the National Science Council in
Taiwan through Contract No.\ NSC100-2112-M-239-001-MY3.
\end{acknowledgments}

%

\bibliographystyle{apsrev4-1}
%

%
%
\end{document}